\documentclass[fleqn,usenatbib]{mnras}

\usepackage{newtxtext,newtxmath}

\usepackage[T1]{fontenc}

\DeclareRobustCommand{\VAN}[3]{#2}
\let\VANthebibliography\thebibliography
\def\thebibliography{\DeclareRobustCommand{\VAN}[3]{##3}\VANthebibliography}

\usepackage{graphicx}	
\usepackage{amsmath}	
\usepackage{comment}

\def\araa{ARA\&A}%
\def\planss{Planet.~Space~Sci.}
\def\bain{Bull.~Astron.~Inst.~Netherlands}

\newcommand\jaavso{JAAVSO}

\def\tm2{$\tau_{\dot{M_2}}$}

\title[A Triple Star Origin For Short-Period Recurrent Novae]{A Triple Star Origin For T Pyx and Other Short-Period Recurrent Novae}

\author[C. Knigge, S. Toonen \& T.C.N. Boekholt]{C. Knigge$^{1,\mathsection}$\thanks{e-mail: c.knigge@soton.ac.uk}, 
S. Toonen$^{2,\mathsection}$\thanks{e-mail: toonen@uva.nl},T.C.N. Boekholt$^{3}$
\\
$^{1}$ Department of Physics and Astronomy, University of Southampton, Southampton, SO17 1BJ, UK\\
$^{2}$ Anton Pannekoek Institute, University of Amsterdam, Postbus 94249, NL-1090 GE Amsterdam, The Netherlands\\
$^{3}$ Rudolf Peierls Centre for Theoretical Physics, Clarendon Laboratory, Parks Road, Oxford, OX1 3PU, UK \\
$^\mathsection$ First and second author contributed equally
}

\date{Accepted XXX. Received YYY; in original form ZZZ}

\pubyear{2021}

\begin{document}
\maketitle

\begin{abstract}
Recurrent novae are star systems in which a massive white dwarf accretes material at such a high rate that it undergoes thermonuclear runaways every 1 - 100 years. They are the only class of novae in which the white dwarf can grow in mass, making some of these systems strong Type Ia supernova progenitor candidates. Almost all known recurrent novae are long-period ($P_{orb} \gtrsim 12~ \mathrm{hrs}$) binary systems in which the requisite mass supply rate can be provided by an evolved (sub-)giant donor star. However, at least two recurrent novae are short-period ($P_{orb} \lesssim 3~\mathrm{hrs}$) binaries in which mass transfer would normally be driven by gravitational radiation at rates 3-4 orders of magnitude smaller than required. Here, we show that the prototype of this class -- T~Pyxidis -- has a distant proper motion companion and therefore likely evolved from a hierarchical triple star system. Triple evolution can naturally produce exotic compact binaries as a result of three-body dynamics, either by Kozai-Lidov eccentricity cycles in dynamically stable systems or via mass-loss-induced dynamical instabilities. By numerically evolving triple progenitors with physically reasonable parameters forward in time, we show explicitly that the inner binary can become so eccentric that mass transfer is triggered at periastron, driving the secondary out of thermal equilibrium. We suggest that short-period recurrent novae likely evolved via this extreme state, explaining their departure from standard binary evolution tracks. 

\end{abstract}

\begin{keywords}
Astrometry and celestial mechanics: proper motions, Stars: novae, cataclysmic variables, Stars: binaries (including multiple): close
\end{keywords}



\section{Introduction}

Most white dwarfs (WDs) in interacting binary systems accrete at a rate below that required for steady Hydrogen burning ($\dot{M}_{H} \simeq 10^{-7}~\mathrm{M_{\odot} yr^{-1}}$;  \citealt{nomoto,fujimoto,wolf, kato-nova}). All such WDs are expected to undergo repeated nova eruptions -- i.e. explosive Hydrogen burning -- when the pressure at the base of non-degenerate accreted layer reaches a critical value. The recurrence time of these eruptions depends primarily on $M_{WD}$, the mass of the WD, and $\dot{M}_{acc}$, the rate at which it accretes  \citep{townsley05}. Most accreting WDs have masses around $M_{WD} \simeq 0.8~\mathrm{M_{\odot}}$ and are found in close binary systems ($P_{orb} \lesssim 12~\mathrm{hr}$) with Roche-lobe-filling $\simeq$main-sequence companions  \citep{knigge06}. Mass transfer is then expected to be driven either by magnetic braking --  in systems with $P_{orb} \gtrsim 3~\mathrm{hr}$, characterized by $\dot{M}_{acc, mb} \simeq 10^{-9}\,-\,10^{-8}~\mathrm{M_{\odot}~yr^{-1}}$ -- or by gravitational radiation -- in systems with $P_{orb} \lesssim 3~\mathrm{hr}$, characterized by $\dot{M}_{acc, gr} \simeq 10^{-11}\,- \,10^{-10}~\mathrm{M_{\odot}~yr^{-1}}$ \citep{knigge11}. Under these conditions, the expected recurrence times are long, $\tau_{rec} \gtrsim 10^{4}~\mathrm{yr}$, and more mass is ejected during eruptions than accreted between eruptions \citep{Yaron05}. In such systems, the WD will therefore never grow to the Chandrasekhar limit, $M_{Ch} \simeq 1.4~\mathrm{M_{\odot}}$.

A small sub-class of {\em recurrent} novae (RNe) erupts much more frequently, $\tau_{rec} \lesssim 100~\mathrm{yr}$ \citep{webbink87,Schaefer10,anu13, Pagnotta14,Darnley14}. This is only possible \citep[e.g.]{Yaron05} if the WD is massive ($M_{WD} \gtrsim 1~\mathrm{M_{\odot}}$) and accretes rapidly ($\dot{M}_{acc, RN} \simeq 10^{-8} \, - \, 10^{-7}~\mathrm{M_{\odot}~yr^{-1}}$). In this regime, the WD {\em is} expected to grow in mass, making RNe viable Type Ia supernova progenitors \citep{livio92,livio18}. Such high accretion rates can be provided by either a massive donor star ($M_2 > M_{WD}$; enabling thermal-timescale mass transfer) or an evolved secondary (enabling accretion from the dense giant wind). Both scenarios \citep{nomoto00, hachisu01} imply large binary systems with long orbital periods ($P_{orb} \gtrsim 6~\mathrm{hr}$). In line with this, most known RNe are characterized by $P_{orb} \gtrsim 12~\mathrm{hrs}$ \citep{Schaefer10}.

The famous RNe T~Pyx has long been a mysterious outlier in this regard. It erupts approximately every $\simeq 30\,\mathrm{yrs}$ \citep{Schaefer10}, even though its orbital period is only $P_{orb} \simeq 1.8~\mathrm{hrs}$ \citep{patterson98, uthas10}. There is no doubt that its outbursts are genuine nova eruptions, nor that its accretion rate between eruptions is extraordinarily high ($\dot{M}_{acc} \simeq 10^{-7}~\mathrm{M_{\odot}~yr^{-1}}$) \citep{Patterson17}. Yet its donor star must be either a very low-mass main-sequence star or even a sub-stellar object ($M_2 \lesssim 0.1~\mathrm{M_{\odot}}$; \citealt{uthas10}). The expected mass-transfer rate along the normal evolution track for such systems is $\dot{M}_{acc, gr}$ \citep{knigge11}, at least 1000 times lower than required by both nova models and observations. Another RN with $P_{orb} \simeq 2.4~\mathrm{hrs}$ -- IM~Normae -- appears to be a near twin of T~Pyx \citep{Schaefer10, Patterson20}. 

A promising way to account for the abnormally high accretion rates in these systems is wind-driven mass transfer \citep{Knigge00}. Since the donor star is strongly irradiated, a powerful outflow is expected to be driven from its surface. Some of this material may directly feed the WD. But even if it escapes, the mass and angular momentum loss associated with this outflow will lead to an increase in the mass-transfer rate through the inner Lagrangian point. This, in turn, may keep the accretion luminosity high enough to sustain the irradiation-driven wind, creating a self-regulating high-$\dot{M}$ state \citep{vanteeseling98, king98}. However, most novae -- recurrent or not -- have clearly evaded this state. The key unsolved question is therefore what makes T~Pyx (and IM~Normae) so special.

\citet{Knigge00} speculated that T~Pyx's current state may have been triggered by residual nuclear burning in the aftermath of a "normal" classical nova eruption. This idea of a classical nova event as the trigger received some observational support from \citet{schaefer10a}, who analysed the physical properties and proper motions of knots in the system's nova shell. Based on this, they argued that T~Pyx must have undergone a classical nova eruption in 1866 $\pm$ 5, whose characteristics were different from those of the recurrent nova outbursts we know of. \footnote{\citet{schaefer10a} also argued that T~Pyx would not erupt again until at least 2225. This prediction did not age well, however, as T~Pyx promptly erupted in the following year.}
However, even if a classical nova event {\em was} the trigger of T~Pyx's current state, the question remains: what makes T~Pyx and IM~Nor special? After all, most classical novae clearly do not trigger high-$\dot{M}_{acc}$ RN states. Is a combination of short orbital period and relatively high WD mass really enough?

One mechanism that can dramatically change the evolutionary path of a close binary system is the gravitational influence of a distant tertiary companion \citep{Toonen16, Too20}. In an effort to test if such companions are common among accreting WD binaries, we have recently carried out a comprehensive search for common proper motion objects in the astrometric data for such sytems provided by ESA's {\em Gaia} mission \citep{gaia16}. T~Pyx is found to have a high-confidence common proper motion companion in this search. Below, we first provide details of the analysis that led to this identification and determine the nature of the proper motion companion. We then present the results of extensive simulations that are designed to explore whether triple evolution can, in fact, produce systems like T~Pyx. We finally return to Schaefer et al's (2010) hypothesis of a classical nova eruption in $\simeq 1866$ as the trigger for T~Pyx's current state and ask whether this can be compatible with a triple evolution scenario.
\section{Finding Common Proper Motion Companions to Accreting White Dwarfs in {\em Gaia} EDR3}
\label{search}

The proper motion companion to T~Pyx was discovered during a comprehensive search for such partners to accreting white dwarfs. Details of the search method and resulting catalogue will be presented in a separate publication, but a brief summary is provided here for completeness.

Our starting point is the catalogue of accreting WDs and candidates previously used in the construction of a volume-limited sample of such systems from {\em Gaia} DR2 \citep{pala20}. This contains $\simeq 8000$~objects, $\simeq 1800$~of which fall within 5\arcsec of a source whose parallax was measured to $\geq 5\sigma$ in the {\em Gaia} Early Data Release~3 (EDR3) catalogue \citep{gaia21}. In order to identify potential proper motion companion to these sources, we largely follow steps developed for the identification of wide binaries in the {\em Gaia} data base \citep{andrews17}. 

First, we require that the separation between physically associated objects should be $s \lesssim 10^{5}$~AU, which corresponds to an orbital period of $\simeq 3\times 10^{7}$~yrs for physically bound systems with a total mass of 1~$M_{\odot}$. This cut-off is based on the largest separations found in the latest version of the Multiple Star Catalog \citep{tokovinin18}. Thus our search radius for companions ranges from $\simeq 2500$\arcsec for the nearest system at $d \simeq 40$~pc to $\simeq 12$\arcsec\ for the most distant system at $d \simeq 8.4$~kpc. Thus we identify all {\em Gaia} sources within this search radius of a given accreting WD "parent" system.

Second, we retain only those sources whose parallaxes agree with that of their parent to better than $5\sigma$.

Third, we check whether the proper motions of these candidate companions are also compatible with those of their parents. The main challenge here is that "compatibility" in this sense is not limited to statistical considerations. More specifically, there are two astrophysical effects that may cause genuine, physically associated objects to have statistically inconsistent proper motions. First, even for bound systems, the orbital motions of the close binary and the tertiary about their center of mass can induce an apparent offset between their proper motions. Second, objects may be (or have been) physically associated even if they are not currently gravitationally bound. Hierarchical triples can evolve in extremely complex ways, e.g. due to von Zeipel-Lidov-Kozai cycles \citep{Von1910, Lid62, Koz62} and/or changes in component masses as a result of stellar evolution. In some cases, this evolution will lead to the ejection of one of the components (typically, but not always, the lowest mass member of the system) \citep{Toonen16, Too20}.

We deal with the first point as follows \citep{andrews17}. We expect bound systems to satisfy $\Delta v \simeq \sqrt{G M_{tot} / s}$, where $M_{tot}$ is the total mass of the system, $s$ is the projected separation between the compact binary and its distant companion, and $\Delta v$ is the difference in the observed projected velocities of these components. This constraint can be turned into an approximate limit on the allowed proper motion difference between gravitationally bound objects. This limit depends on their projected separation (which is known for our candidates) and the total system mass (for which we conservatively adopt $M_{tot} = 10~M_{\odot}$). We thus require that physically associated objects should either have statistically compatible proper motions (at the 5$\sigma$ level or better), or -- if the proper motions are incompatible -- that the difference between these proper motions should be compatible with the limit expected for bound systems (again to within 5$\sigma$). 

As just noted above, this still leaves open the possibility that some physically associated, but unbound, systems will be missed. We do not explicitly expand our search to target such systems, but expect that many of them will be included in our catalogue regardless. This is because taking $M_{tot} \simeq 10~M_{\odot}$ and allowing up to $5\sigma$ disagreements makes our filtering process quite conservative, in the sense that false positives should be considerably more likely than false negatives. Indeed, as discussed below and in the main text, T~Pyx and its companion exhibit formally incompatible proper motions.

This selection yields a sample of $\simeq 300$ systems with one or more viable companions based on parallax and proper motion. As a final cut, we focus on objects with a {\em single} candidate companion, since multiple viable matches to a single source will typically be false positives in regions of extremely high stellar density. This yields a final sample of $\simeq 180$ systems, which includes T~Pyx. 

Except for the adoption of slightly different parameters in our cuts, our sample is selected in an extremely similar way to that used in the construction of the recently released {\em Gaia} EDR3 million-binary catalogue (MBC) \citep{elbadry21}. In line with this, we can recover all of the accreting WDs with proper motion companions within 100,000~AU in the MBC. However, since the MBC is limited to systems closer than $d \lesssim 1$~kpc, it does not include T~Pyx.

\section{T~Pyx's Proper Motion Companion}

\begin{figure}
\includegraphics[width=8.5cm]{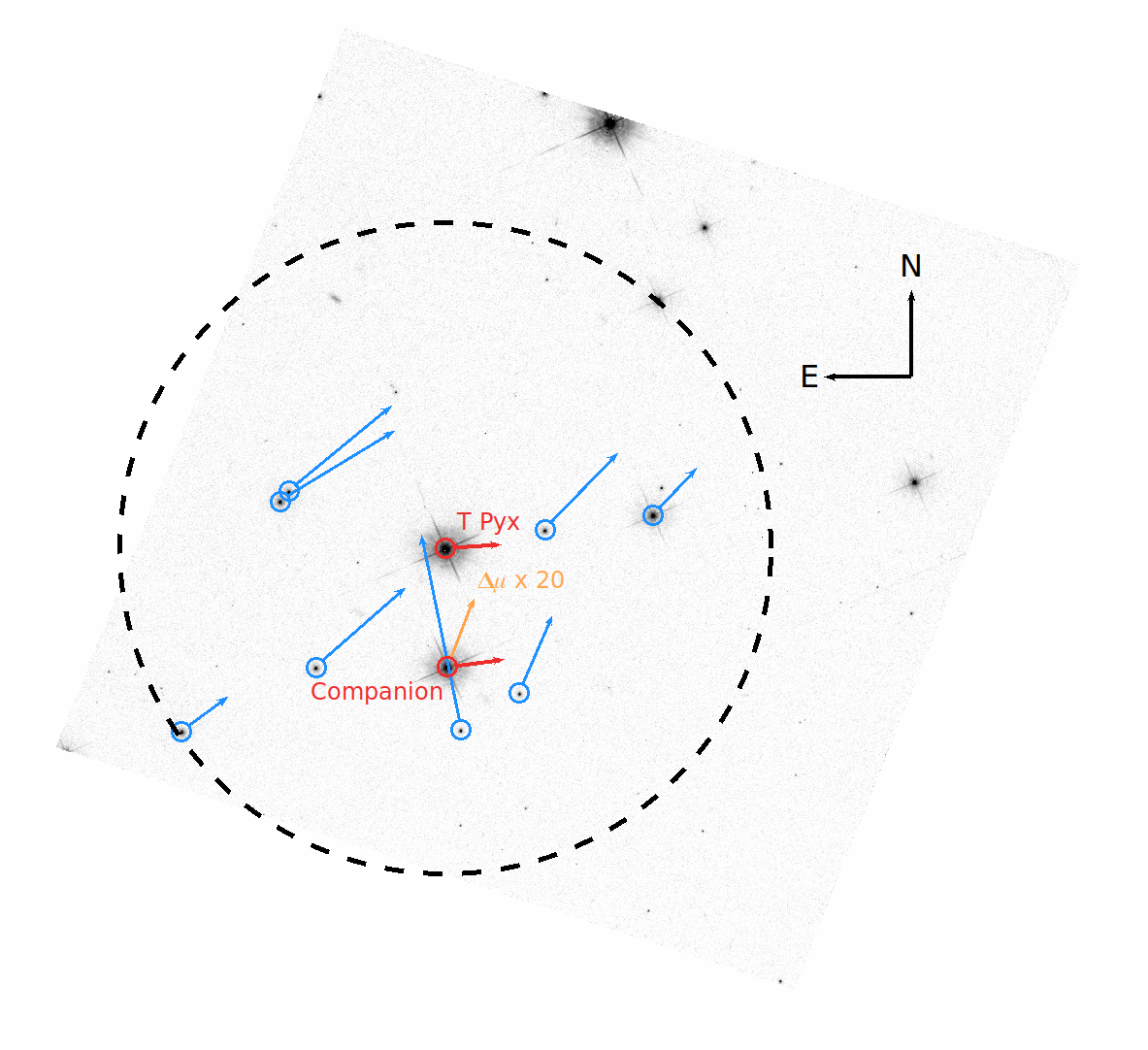} 
\caption{{\bf Hubble Space Telescope / Wide Field Camera 3 image of the field around T~Pyx in the optical F547M filter.} The dashed black circle contains the area in which we searched for {\em Gaia} proper motion companions. All {\em Gaia} sources with good parallax and proper motion measurements in this area are marked with circles, and their proper motion vectors are shown by arrows. The two sources shown in red are T~Pyx and its companion. The orange arrows shows the vector difference between their proper motion, magnified by a factor 20. }\label{fig:image}
\end{figure}

The offset between T~Pyx and its companion is 12.4\arcsec, corresponding to $7.7 \times 10^6$~R$_{\odot}$ at the Gaia EDR3 distance of 2.9~kpc towards the system \citep{gaia21}.
As noted above, T~Pyx is included in our final sample of CVs with candidate proper motion companions.  Based on extensive Monte Carlo simulations described in the Appendix, the single-trial false-alarm probability is $\lesssim 1$\%, and a physical association with the companion is favoured even if we ignore the system's unusual nature and 
evaluate the match as a random trial among many (i.e. one for each object in our master list of accreting WDs).

Figure~\ref{fig:image} shows a high-resolution optical image of the field around T~Pyx obtained by the {\em Hubble Space Telescope}. Our search radius around the system is shown by the dashed line. All {\em Gaia} sources with reliable data inside this radius are highlighted with open circles and shown with their proper motion vectors. The suggested companion clearly stands out as the only object whose proper motion is compatible with that of T~Pyx.

\begin{figure*}
\centering
\includegraphics[width=17cm]{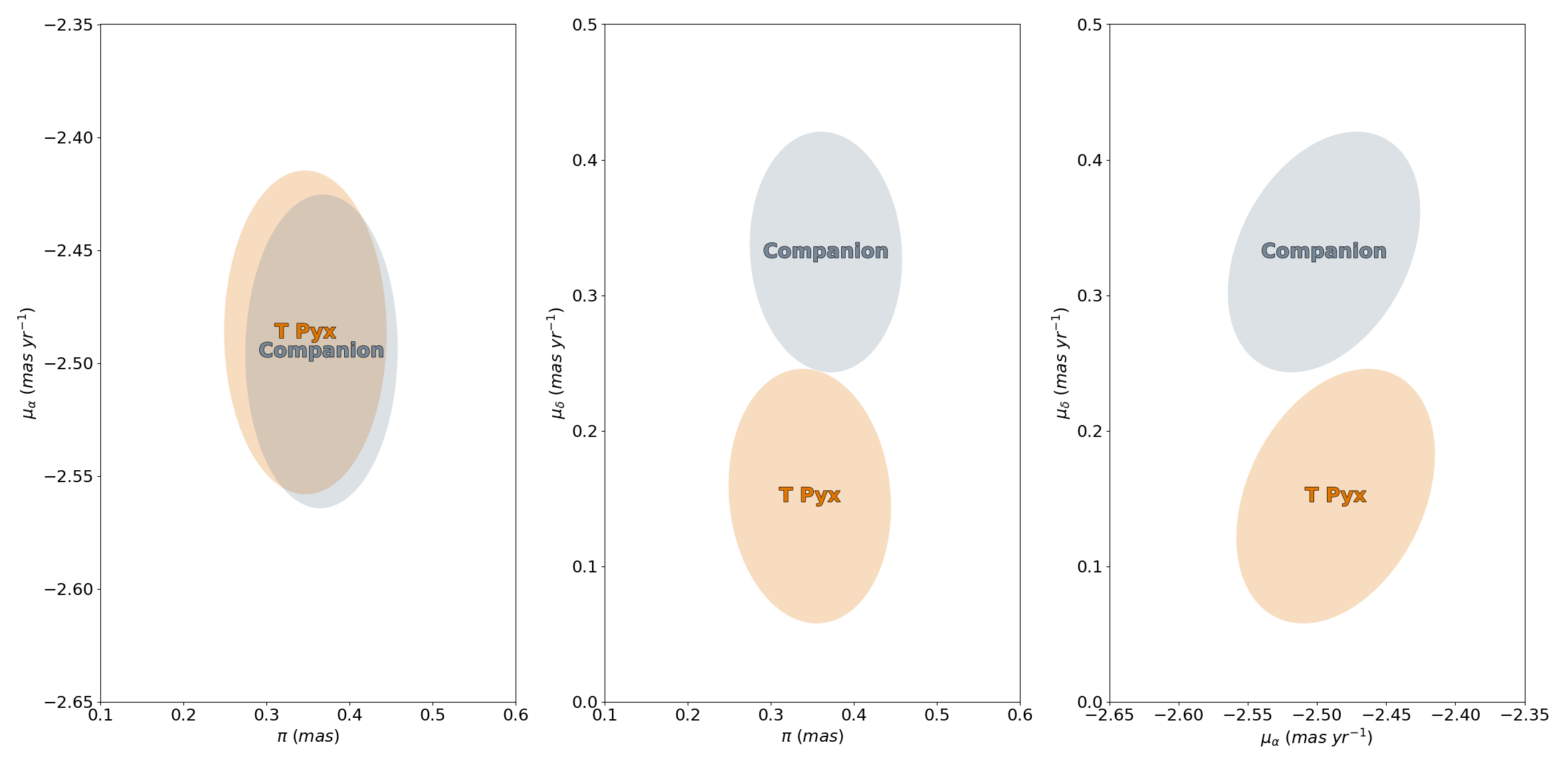} 
\caption{\textbf{Parallax and proper motion diagrams for T~Pyx and its companion.} Each panel shows one pairwise combination of $\pi$, $\mu_{\alpha}$ and $\mu_{\delta}$. The ellipses correspond to 99.7\% confidence regions around the measured values for each source, including the correlations between the parameter.}\label{fig:parapm}
\end{figure*}

Figure~\ref{fig:parapm} shows the locations of T~Pyx and its putative companion in parallax and proper motion space. Crucially, the {\em Gaia }parallaxes -- and hence distances -- of the two objects are almost identical and well within their respective uncertainties. Their proper motions are also similar, although they formally disagree at the roughly $3\sigma$ level. As discussed further below, this difference may well be real and can be understood within a triple scenario. For example, the mass loss associated with the transition of the primary from a main-sequence star to a WD can naturally lead to the dissolution of such systems.

\begin{figure}
\centering
\includegraphics[width=8.5cm,trim=0cm 0cm 0cm 0.5cm,clip]{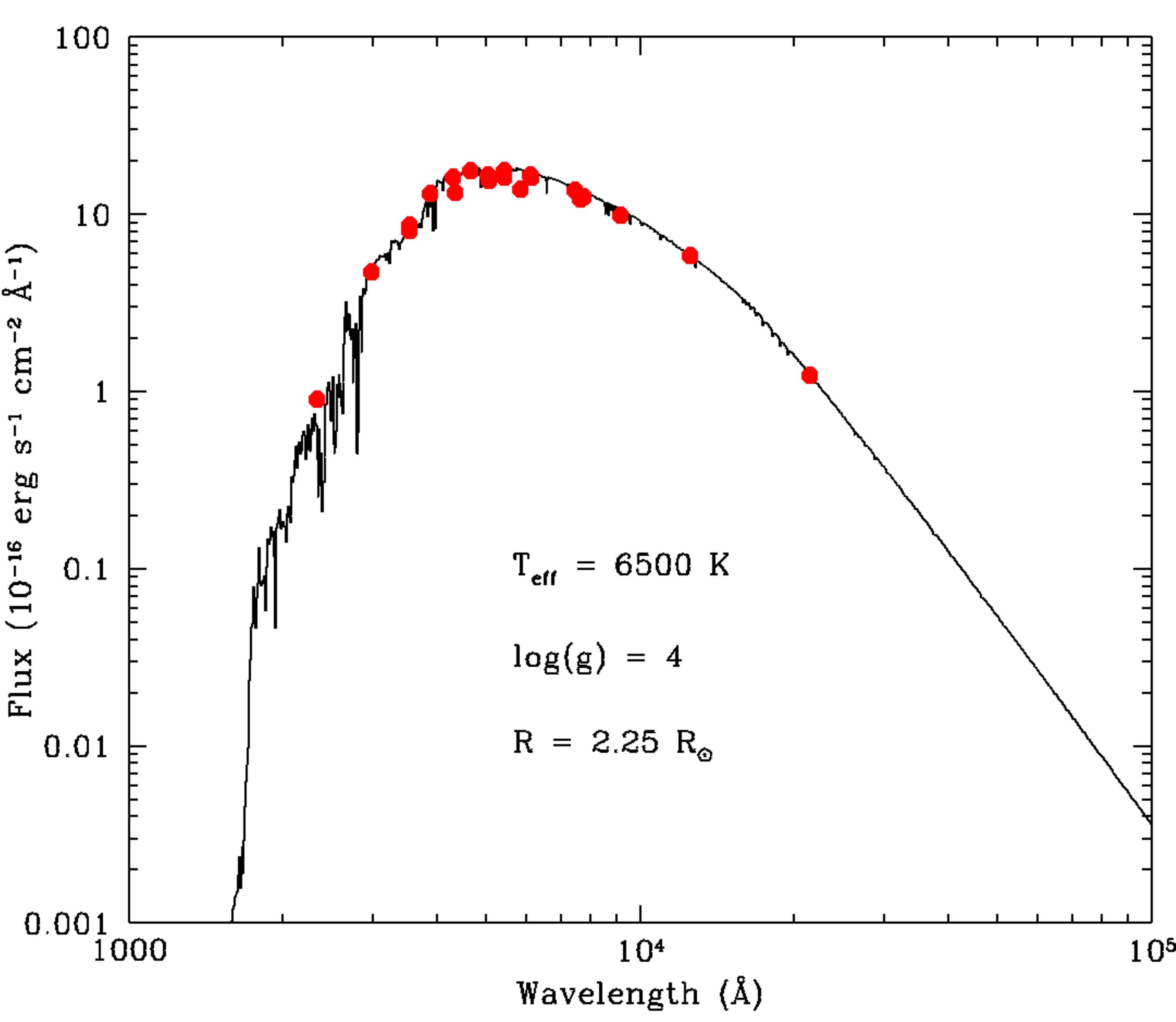} 
\caption{\textbf{The spectral energy distribution of the companion.} The red points correspond to photometric measurements spanning the near-ultraviolet to mid-infrared region \citep{vista1,vista2,skymapper1,skymapper2,xmmom, apass1, apass2, gaia21, gaiaphotom,2mass,wise}. The black line shows a simple model stellar atmosphere \citep{castelli03} that matches the observed SED.}
\label{fig:sed}
\end{figure}

In order to shed light on the physical nature of the companion, we have collated photometric measurements from a variety of sources, covering the near-ultraviolet through mid-infrared bands. The resulting spectral energy distribution (SED) is shown in Figure~\ref{fig:sed}, along with the spectrum predicted by a stellar atmosphere model with effective temperature $T_{eff} \simeq 6500$~K and surface gravity $\log{g} = 4$. These parameters imply a radius of $R_{3} \simeq 2.25~\mathrm{R_{\odot}}$ for the erstwhile tertiary. 

Additional information can be obtained by placing the system on a colour-magnitude diagram and comparing its location to theoretical stellar evolution tracks. Such a comparison is shown in Figure~\ref{fig:cmd} and shows that the object is likely to be a $M_{3} \simeq 1.5~\mathrm{M_{\odot}}$ sub-giant. This is an evolutionary phase during which the star moves from the stellar MS towards the red giant branch after exhausting the Hydrogen fuel in its core. Since the sub-giant phase is relatively short-lived, this allows us to estimate the age of the star -- and thus probably of T~Pyx as well. This turns out to be $\simeq 2.2$~Gyr.

\section{Simulations: How Triple Evolution Can Produce A System like T~Pyx}

Having established the properties of the proper motion companion, as well as its likely physical association with T~Pyx, we now address whether and how such a distant companion can account for the abnormal characteristics of this system (relative to other RNe and accreting WDs). Hierarchical triples -- in which two stars orbit around each other, while a distant tertiary orbits around the inner binary -- are actually quite abundant. Amongst solar-type (primary) stars, for every three binaries, there is roughly one triple \citep{Tok14b, Moe17}. This multiplicity fraction increases with stellar mass: O- \& B-type stars are rarely found as singles or in "pure" binary systems, but are typically members of triples and higher-order multiples \citep{Rem11,San12,Moe17}. 

Stellar interactions (i.e. mass transfer episodes) are considerably more prevalent in triples than in binaries, e.g. by a factor $\sim$2-3 among solar-mass stars\footnote{The fraction of low- and intermediate mass triples that will experience mass transfer in their evolution is 2-3 times larger than for binaries in the same mass range \citep{Too20}.}. Triple evolution is thus quite a natural mechanism for the formation of close WD binaries. In fact, at least $\simeq 7\%$ (3/42) of the accreting WD binary systems within 150~pc are known or suspected to have a tertiary companion \citep{pala20}.

In order to account for systems like T~Pyx, we specifically consider the three evolutionary channels sketched in Fig.\,\ref{fig:cartoon}. We have explored each of these channels quantitatively by carrying out simulations with the triple evolution code \texttt{TRES}\footnote{\texttt{TRES} is publicly available through the github repository of the Astrophysics MUltipurpose Software Environment \citep{Por18} (AMUSE).}, which combines three-body dynamics with stellar evolution and dissipative processes at run-time \citep{Toonen16}. It is the coupling between these processes -- which include wind-driven mass loss, tidal effects and Lidov-Kozai cycles -- that drive stellar interactions in efficient ways  \citep{Kis98,Fab07, Perets12, Shappee13, Michaely14, Too20}. 

Here, we give only a brief overview of each channel, along with the main results of our simulations; in Appendix D, we additionally provide a detailed description for one illustrative simulation per channel. All channels start with three stars on the zero-age main sequence, in a configuration consisting of a primary and secondary in a binary, which is orbited by a tertiary in a  wider orbit. The channels start to diverge once the primary leaves the main sequence and starts to ascend the giant branch.

\begin{enumerate}
    \item \emph{The Kozai channel:} 
    As the primary star evolves, it loses mass via a stellar wind before settling down on the WD sequence. 
    The inner and outer orbits are wide enough to avoid Roche-lobe overflow, and they widen further due to adiabatic wind mass losses. Since the fractional mass loss is larger for the inner binary compared to the outer "binary", the inner orbit widens more relative to the outer one, and the hierarchy of the system decreases. As the two orbits approach one another, 
    strong gravitational perturbations between the inner and outer orbit ensue. 
    Lidov-Kozai cycles \citep{Von1910,Lid62,Koz62} 
    and higher-order dynamical effects \citep{Nao16} become increasingly important, and enhance or even initiate eccentricity variations in the inner binary. Eventually these lead to Roche-lobe overflow (described in detail below), and the formation of a cataclysmic variable (CV).
    
    \item \emph{The dynamical instability channel:} Here, as in the Kozai channel, stellar wind mass losses reduce the hierarchy of the system. However, in this channel, the effect is so strong that the system becomes dynamically unstable \citep{Kis94,Ibe99,Perets12}. The resulting orbital evolution (step 3-4 in Fig.\,\ref{fig:cartoon}) is chaotic \citep{Perets12,He18,Too21,Ham21} and results either in strongly varying inner and outer orbits, or in a complete loss of hierarchy (reminiscent of binary-single encounters in globular clusters). It triggers close encounters, collisions, ejections, and the formation of compact and interacting binaries (see below for details). As the instability is generally driven by stellar wind mass loss, this channel naturally leads to triples with WD components. 
    
    \item \emph{The common envelope channel:} The third channel we consider is essentially the canonical (binary) evolution channel for the formation of accreting WD systems, but in the presence of a distant tertiary companion. Here, the inner binary experiences a common-envelope (CE) phase during the primary's ascent of the giant branch. 
    The CE-phase is instigated by the evolutionary expansion of the primary star, or the inner binary may be driven to it by three-body dynamics. 
    As a consequence of the CE-phase, the inner orbit  shrinks, and the WD progenitor is stripped of its envelope. The resulting mass loss may lead to the dissolution of the triple system, i.e. the ejection of the tertiary. The compact inner binary, now comprising a WD and a MS star, becomes more compact over time due to magnetic braking and gravitational wave emission. Eventually, this leads to the onset of mass transfer and the formation of a CV \citep{knigge11}. 
\end{enumerate}

These channels can form extreme systems -- like T~Pyx -- because the inner binary in a triple is far more likely to be significantly eccentric at the onset of mass transfer than a canonical-channel zero-age CV.  In both binaries and triples, tides tend to circularize and synchronize the stellar motions. However, in triples, this is counteracted by three-body dynamics, which can drive eccentricity variations and growth. As a result, isolated binaries are expected to be circular at the onset of mass transfer  \citep{Hur02}, but this is not the case for triples \citep{Too20}. Roche-lobe overflow will then initially occur only near periastron. The resulting mass transfer rate is high, strongly orbital phase-dependent and can evolve much more rapidly than in canonical CVs \citep{lajoie,davis13,nelemans16}. These conditions may provide the necessary trigger for T~Pyx-like behaviour  \citep{Knigge00}.

Having established that T-Pyx-like systems are a {\em possible} product of each channel, we next consider if their formation is {\em likely}. To this end, we have performed a triple population synthesis study with \texttt{TRES}. Details are provided in Appendix~B; here we provide only a brief overview of the main results. 
We estimate a Galactic birth rate of $10^{-4}$ per year for the Kozai channel, $10^{-5}-10^{-4}$ per year for the dynamical instability channel, and several $10^{-3}$ per year for the common envelope channel. For comparison, the best observational estimate \citep{pala20} of the overall CV space density is $\simeq 5 \times 10^{-5}$~pc$^{-3}$. For a characteristic observable CV lifetime of 5~Gyr, this translates to a birthrate of $\simeq 5 \times 10^{-4}$ per year \citep{elbadry21}. Thus triple evolution can easily produce the relatively small number extreme CVs we are considering here. It might even be an important formation channel for CVs overall.  

Our simulations show that the tertiary is typically at a large distance from the CV (Extended Data Figure~\ref{fig:orbit*3}), which is consistent with what we observe for T~Pyx. The median distance in the Kozai and dynamical instability channels is
$(1-3)\times 10^6~R_{\odot}$ and 
$(3-4)\times 10^7~R_{\odot}$, respectively.
In the CE channel, the median distance is $8.0\times10^4 - 5.3\times 10^5~ R_{\odot}$ at the onset of the CE-phase. Depending on the timescale and morphology of the mass ejection, the tertiary will likely become unbound due to the CE mass loss  \citep{Mic19,Igo20}. Ionisation is also possible for weakly bound \citep[$a_3 \gtrsim  10^6~R_{\odot}$, ][]{Too17} tertiaries in the other two channels, if the timescale of wind mass loss is significantly shorter than one orbital period \citep{Had66,Alc86,Ver11,Too17}.

The simulations also show that all WDs in the Kozai and dynamical instability channels are carbon-oxygen (or oxygen-neon) WDs in line with CV observations. Only in the CE channel do we find accreting helium WDs. The lack of helium WDs in the Kozai and dynamical instability channels is an inherent characteristic. These channels require significant mass loss from the system, which is strongest on the asymptotic giant branch (AGB) -- a phase only reached by primaries after the Helium flash.

Importantly, our simulations show that orbits can be either circular ($e_{\rm in} \lesssim 10^{-3}$), or significantly eccentric ($e_{\rm in} > 0.1$) at the onset of a mass transfer phase. As discussed in Appendix C, this holds for the CE channel, but also for the start of the CV phase in the Kozai and dynamical instability channels. For the latter, we distinguish three scenarios (steps 3-5 in Fig.\,\ref{fig:cartoon}):

\begin{figure}
\centering
\includegraphics[width=8.5cm,trim=0cm 0cm 0cm 0.5cm,clip]{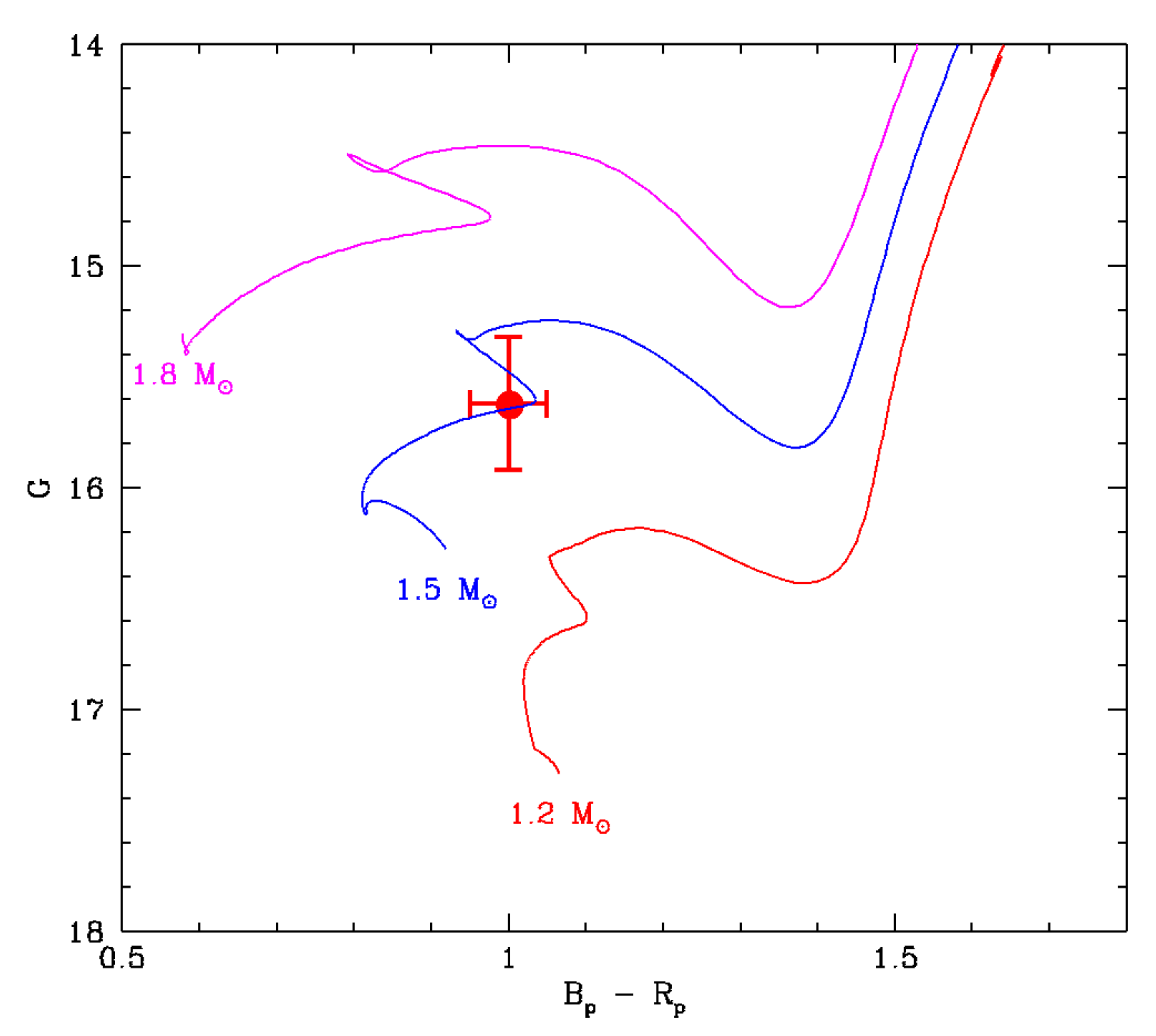} 
\caption{\textbf{Optical colour-magnitude diagram comparing the observed properties of the  companion (red circle with error bars) to theoretical evolution tracks for stars with different masses.} \citep{mist0,mist1}
This comparison suggests that the companion is a $\simeq 1.5~M_{\odot}$ star that is about to transition from the terminal-age main sequence to the sub-giant branch. These are a relatively short-lived evolutionary phases, providing an age estimate of $\simeq 2.2$~Gyr for the star.
The observational error bars shown here are approximate, but conservative; they are dominated by uncertainties in the distance and the extinction/reddening towards the source.}
\label{fig:cmd}
\end{figure}

\begin{enumerate}

\item On the extreme end (during the most intense dynamical interactions in step 3 of the dynamical instability channel), strong gravitational perturbations between the stars lead to nearly parabolic orbits and stellar collisions. These occur typically between the two stars of the initial inner binary \citep{Too21}. Hydrodynamical simulations show that the collision between a WD and MS star lead to the tidal disruption of the MS star or to a significant stellar disruption (in the case of massive MS stars) \citep{Sha86,Reg87,Sok87,Roz89}.

\item For less extreme eccentricities, a head-on collision is avoided. Instead gas is tidally stripped from a star and accreted on its companion  at or close to periastron passage \citep{Lay98,Reg05,Sep07,Sep09,Sep10,Chu09,Laj11,Dav13,Van16,Dos16,Dos16b, Ham19}. The stars do not physically touch, but mass is transferred periodically. This process, also known as eccentric mass transfer, leads, for example, to X-ray flares in high-mass X-ray binaries \citep{Lay98}. Regarding T~Pyx specifically, the thermal timescale of a 0.1M$_{\odot}$ (0.2M$_{\odot}$) MS star is on the order of 2.5~Gyr (1~Gyr) \citep{Too12}, much longer than the orbital periods of the inner binary in the Kozai and dynamical instability channels ($\simeq 10^1-10^2$~yr and $\simeq(5-8)\times 10^3$~yr, respectively. Thus a T-Pyx-like donor star will easily be driven out of thermal equilibrium by the periodic gas stripping. As the orbit slowly circularizes over time, the intense bursts of mass accretion onto the WD can then kick-start an irradiation-induced wind-driven mass-transfer phase \citep{Knigge00,Too14b}.

\item For even less extreme eccentricities, a compact binary forms. This occurs very efficiently through the interplay of three-body dynamics with tidal friction \citep{Maz79, Kis98, Fab07, Liu15, Bat18}. As the distant tertiary star induces long term eccentricity excitations, there is strong dissipation during the periastron passages in the form of tides, gravitational wave emission, or magnetic braking. For most of the orbit when the stars are at large separations from one another, dissipation is negligible, but during the periastron passages strong tidal torques decay the orbit. 
Eventually, the inner orbit becomes so compact, that the tidal torques quench any remaining Lidov-Kozai cycles. Another possibility is that the freeze-out does not occur gradually, but that the inner orbit ends up in this regime after one strong encounter with the tertiary star, which is possible for unstable or mildly-hierarchical systems \citep{Cuk04,Ant12,Kat12,Luo16, Gri18,Liu18,Rod18,Bha21}. 
Afterwards, the compact binary will come in contact and become a CV through magnetic braking and gravitational wave emission. 
\end{enumerate}

\begin{figure*}
\centering
\includegraphics[width=16.5cm,trim=0cm 0cm 0cm 0.5cm,clip]{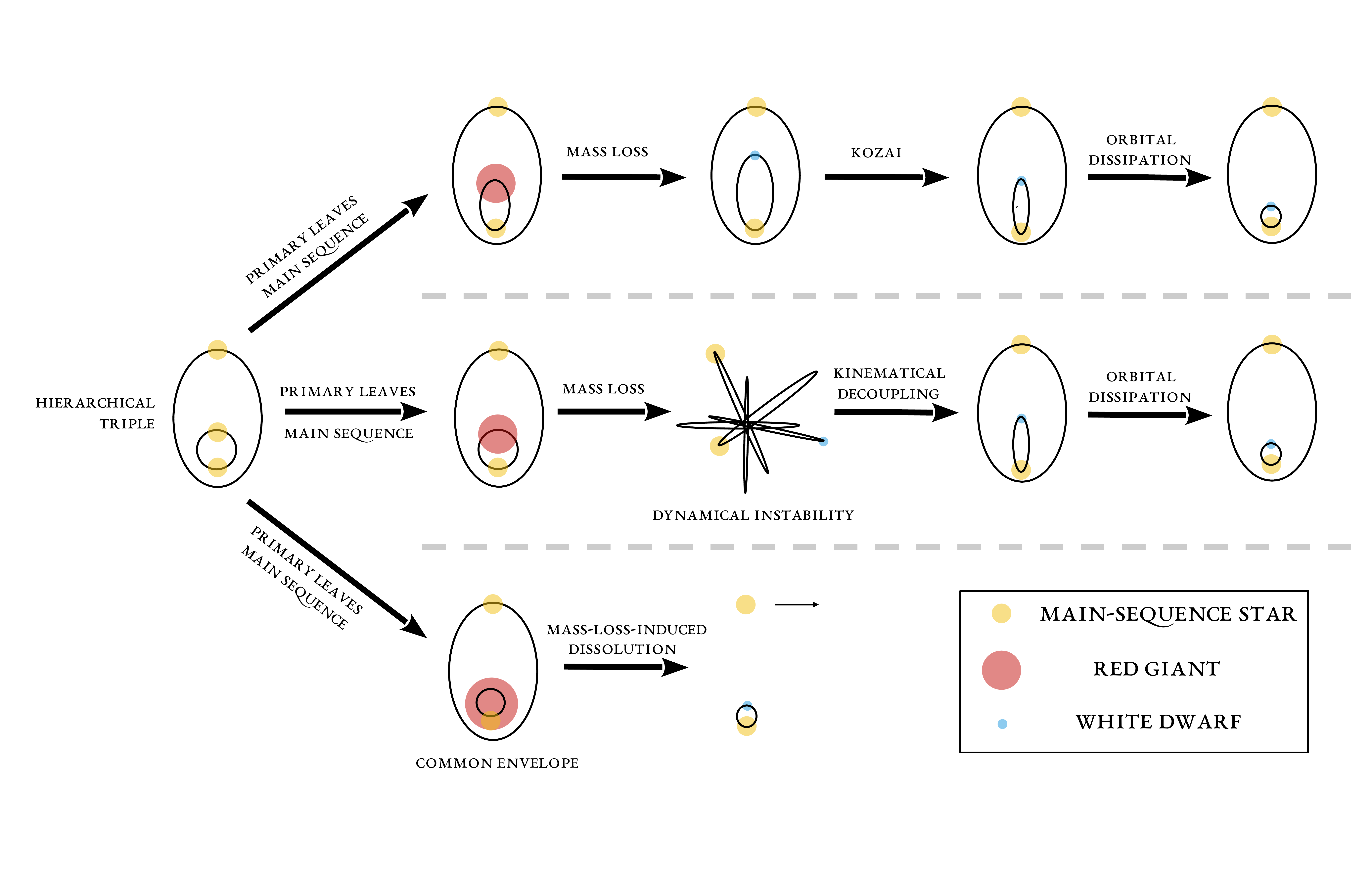} 
\caption{\textbf{Three triple-star evolution channels that can produce accreting white dwarfs with low-mass main-sequence donor stars and distant proper motion companions.} See text for details.}
\label{fig:cartoon}
\end{figure*}

\section{Discussion and Conclusions}

We have shown that the short-period RN T~Pyx has a proper motion companion in
{\em Gaia}~EDR3 and therefore probably evolved in a hierarchical triple. Based on extensive numerical simulations, we have also found that triple evolution can naturally produce accreting WDs in compact binaries with distant (bound or unbound) proper motion companions, at rates that can easily account for a system like T~Pyx. 

It is interesting to ask if a triple evolution scenario for T~Pyx might be compatible with Schaefer et al.'s (2010) hypothesis that T~Pyx's RN state was triggered by a classical nova eruption in $\simeq$1866. On the one hand, classical novae are generally thought to be associated with relatively long, quasi-steady periods of mass accumulation at low $\dot{M}_{acc}$. This is hard to reconcile with a triple scenario in which -- as we have shown -- the current high-$\dot{M}_{acc}$ RN state is likely associated with a highly eccentric inner binary coming into contact. On the other hand, the notion that classical novae are associated with long periods of accretion at low $\dot{M}_{acc}$ is largely based on theoretical simulations in which the accreting and erupting WD has reached a quasi-steady state with roughly constant recurrence times. As demonstrated explicitly by \citet{epelstain}, such a state is only reached after hundreds of eruptions. So is it possible that the putative 1866 eruption may actually have been (one of) the {\em first} eruption(s) in a system brought into contact by triple evolution? This question is unfortunately out of reach at the moment, since it would require simulations that self-consistently include three-body dynamics, (eccentric) mass-transfer and nova evolution. We are therefore agnostic on this issue. 

In any case, if triple evolution is indeed the explanation for T~Pyx's extreme and unsual properties, its only known twin -- IM~Nor \citep{Patterson20} -- may also be expected to have a proper motion component. Unfortunately, IM~Nor does not have a statistically meaningful parallax estimate in {\em Gaia}~EDR3 ($\pi = -0.48 \pm 0.33$~mas). However, this may change in future data releases.

\section*{Acknowledgments}

This project was supported by funds from the European Research Council (ERC) under the European Union’s Horizon 2020 research and innovation program under grant agreement No 638435 (GalNUC). 

\section*{Data Availability}

The Gaia EDR3 data base is publicly available from the {\em Gaia} Archive (https://gea.esac.esa.int/archive/). The Hubble Space Telescope data used in the construction of Figure~1 is available via the MAST archive ({https://archive.stsci.edu}). The theoretical models used in Figures~3 and 4 are available for download from wwwuser.oats.inaf.it/castelli/grids/gridp00k2odfnew/fp00k2tab.html and waps.cfa.harvard.edu/MIST/index.html, respectively. All but one of the photometric data sets used to construct the spectral energy distribution (SED) in Figure~3 is available from the {\em Vizier} service provided by the CDS\footnote{https://vizier.u-strasbg.fr/viz-bin/VizieR}. The one exception is the APASS optical photometry, which is available online from the AAVSO\footnote{https://www.aavso.org/apass}. In constructing the SED, we also used the bandpass properties provided by the Spanish Virtual Observatory\footnote{http://svo2.cab.inta-csic.es/theory/fps/}.

\bibliographystyle{mnras}
\bibliography{sample}

\begin{thebibliography}{}
\makeatletter
\relax
\def\mn@urlcharsother{\let\do\@makeother \do\$\do\&\do\#\do\^\do\_\do\%\do\~}
\def\mn@doi{\begingroup\mn@urlcharsother \@ifnextchar [ {\mn@doi@}
  {\mn@doi@[]}}
\def\mn@doi@[#1]#2{\def\@tempa{#1}\ifx\@tempa\@empty \href
  {http://dx.doi.org/#2} {doi:#2}\else \href {http://dx.doi.org/#2} {#1}\fi
  \endgroup}
\def\mn@eprint#1#2{\mn@eprint@#1:#2::\@nil}
\def\mn@eprint@arXiv#1{\href {http://arxiv.org/abs/#1} {{\tt arXiv:#1}}}
\def\mn@eprint@dblp#1{\href {http://dblp.uni-trier.de/rec/bibtex/#1.xml}
  {dblp:#1}}
\def\mn@eprint@#1:#2:#3:#4\@nil{\def\@tempa {#1}\def\@tempb {#2}\def\@tempc
  {#3}\ifx \@tempc \@empty \let \@tempc \@tempb \let \@tempb \@tempa \fi \ifx
  \@tempb \@empty \def\@tempb {arXiv}\fi \@ifundefined
  {mn@eprint@\@tempb}{\@tempb:\@tempc}{\expandafter \expandafter \csname
  mn@eprint@\@tempb\endcsname \expandafter{\@tempc}}}

\bibitem[\protect\citeauthoryear{{Aarseth} \& {Mardling}}{{Aarseth} \&
  {Mardling}}{2001}]{Aar01}
{Aarseth} S.~J.,  {Mardling} R.~A.,  2001, in {Podsiadlowski} P.,  {Rappaport}
  S.,  {King} A.~R.,  {D'Antona} F.,   {Burderi} L.,  eds,  Astronomical
  Society of the Pacific Conference Series Vol. 229, Evolution of Binary and
  Multiple Star Systems. p.~77 (\mn@eprint {} {astro-ph/0011514})

\bibitem[\protect\citeauthoryear{{Alcock}, {Fristrom}  \& {Siegelman}}{{Alcock}
  et~al.}{1986}]{Alc86}
{Alcock} C.,  {Fristrom} C.~C.,   {Siegelman} R.,  1986, \mn@doi [\apj]
  {10.1086/164005}, \href
  {https://ui.adsabs.harvard.edu/abs/1986ApJ...302..462A} {302, 462}

\bibitem[\protect\citeauthoryear{{Andrews}, {Chanam{\'e}}  \&
  {Ag{\"u}eros}}{{Andrews} et~al.}{2017}]{andrews17}
{Andrews} J.~J.,  {Chanam{\'e}} J.,   {Ag{\"u}eros} M.~A.,  2017, \mn@doi
  [\mnras] {10.1093/mnras/stx2000}, \href
  {https://ui.adsabs.harvard.edu/abs/2017MNRAS.472..675A} {472, 675}

\bibitem[\protect\citeauthoryear{{Antognini}}{{Antognini}}{2015}]{Ant15}
{Antognini} J.~M.~O.,  2015, \mn@doi [\mnras] {10.1093/mnras/stv1552}, \href
  {https://ui.adsabs.harvard.edu/abs/2015MNRAS.452.3610A} {452, 3610}

\bibitem[\protect\citeauthoryear{{Antonini} \& {Perets}}{{Antonini} \&
  {Perets}}{2012}]{Ant12}
{Antonini} F.,  {Perets} H.~B.,  2012, \mn@doi [\apj]
  {10.1088/0004-637X/757/1/27}, \href
  {https://ui.adsabs.harvard.edu/abs/2012ApJ...757...27A} {757, 27}

\bibitem[\protect\citeauthoryear{{Anupama}}{{Anupama}}{2013}]{anu13}
{Anupama} G.~C.,  2013, in {Di Stefano} R.,  {Orio} M.,   {Moe} M.,  eds, ~
  Vol. 281, Binary Paths to Type Ia Supernovae Explosions. pp 154--161,
  \mn@doi{10.1017/S1743921312014913}

\bibitem[\protect\citeauthoryear{{Bataille}, {Libert}  \& {Correia}}{{Bataille}
  et~al.}{2018}]{Bat18}
{Bataille} M.,  {Libert} A.-S.,   {Correia} A.~C.~M.,  2018, \mn@doi [\mnras]
  {10.1093/mnras/sty1758}, \href
  {http://adsabs.harvard.edu/abs/2018MNRAS.479.4749B} {479, 4749}

\bibitem[\protect\citeauthoryear{{Bhaskar}, {Li}, {Hadden}, {Payne}  \&
  {Holman}}{{Bhaskar} et~al.}{2021}]{Bha21}
{Bhaskar} H.,  {Li} G.,  {Hadden} S.,  {Payne} M.~J.,   {Holman} M.~J.,  2021,
  \mn@doi [\aj] {10.3847/1538-3881/abcbfc}, \href
  {https://ui.adsabs.harvard.edu/abs/2021AJ....161...48B} {161, 48}

\bibitem[\protect\citeauthoryear{{Blaauw}}{{Blaauw}}{1961}]{Bla61}
{Blaauw} A.,  1961, \bain, \href
  {https://ui.adsabs.harvard.edu/abs/1961BAN....15..265B} {15, 265}

\bibitem[\protect\citeauthoryear{{Boekholt} \& {Correia}}{{Boekholt} \&
  {Correia}}{2022}]{TBTIDY}
{Boekholt} T.~C.~N.,  {Correia} A. C.~M.,  2022, in preparation

\bibitem[\protect\citeauthoryear{{Camacho}, {Torres}, {Garc{\'\i}a-Berro},
  {Zorotovic}, {Schreiber}, {Rebassa-Mansergas}, {Nebot G{\'o}mez-Mor{\'a}n}
  \& {G{\"a}nsicke}}{{Camacho} et~al.}{2014}]{Cam14}
{Camacho} J.,  {Torres} S.,  {Garc{\'\i}a-Berro} E.,  {Zorotovic} M.,
  {Schreiber} M.~R.,  {Rebassa-Mansergas} A.,  {Nebot G{\'o}mez-Mor{\'a}n} A.,
   {G{\"a}nsicke} B.~T.,  2014, \mn@doi [\aap] {10.1051/0004-6361/201323052},
  \href {https://ui.adsabs.harvard.edu/#abs/2014A&A...566A..86C} {566, A86}

\bibitem[\protect\citeauthoryear{{Casagrande}, {Wolf}, {Mackey}, {Nordlander},
  {Yong}  \& {Bessell}}{{Casagrande} et~al.}{2019}]{skymapper2}
{Casagrande} L.,  {Wolf} C.,  {Mackey} A.~D.,  {Nordlander} T.,  {Yong} D.,
  {Bessell} M.,  2019, \mn@doi [\mnras] {10.1093/mnras/sty2878}, \href
  {https://ui.adsabs.harvard.edu/abs/2019MNRAS.482.2770C} {482, 2770}

\bibitem[\protect\citeauthoryear{{Castelli} \& {Kurucz}}{{Castelli} \&
  {Kurucz}}{2003}]{castelli03}
{Castelli} F.,  {Kurucz} R.~L.,  2003, in {Piskunov} N.,  {Weiss} W.~W.,
  {Gray} D.~F.,  eds, ~ Vol. 210, Modelling of Stellar Atmospheres. p.~A20
  (\mn@eprint {arXiv} {astro-ph/0405087})

\bibitem[\protect\citeauthoryear{{Choi}, {Dotter}, {Conroy}, {Cantiello},
  {Paxton}  \& {Johnson}}{{Choi} et~al.}{2016}]{mist1}
{Choi} J.,  {Dotter} A.,  {Conroy} C.,  {Cantiello} M.,  {Paxton} B.,
  {Johnson} B.~D.,  2016, \mn@doi [\apj] {10.3847/0004-637X/823/2/102}, \href
  {https://ui.adsabs.harvard.edu/abs/2016ApJ...823..102C} {823, 102}

\bibitem[\protect\citeauthoryear{{Church}, {Dischler}, {Davies}, {Tout},
  {Adams}  \& {Beer}}{{Church} et~al.}{2009}]{Chu09}
{Church} R.~P.,  {Dischler} J.,  {Davies} M.~B.,  {Tout} C.~A.,  {Adams} T.,
  {Beer} M.~E.,  2009, \mn@doi [\mnras] {10.1111/j.1365-2966.2009.14619.x},
  \href {https://ui.adsabs.harvard.edu/abs/2009MNRAS.395.1127C} {395, 1127}

\bibitem[\protect\citeauthoryear{{{\'C}uk} \& {Burns}}{{{\'C}uk} \&
  {Burns}}{2004}]{Cuk04}
{{\'C}uk} M.,  {Burns} J.~A.,  2004, \mn@doi [\aj] {10.1086/424937}, \href
  {https://ui.adsabs.harvard.edu/abs/2004AJ....128.2518C} {128, 2518}

\bibitem[\protect\citeauthoryear{{Cutri} \& {et al.}}{{Cutri} \& {et
  al.}}{2012}]{wise}
{Cutri} R.~M.,  {et al.} 2012, VizieR Online Data Catalog, \href
  {https://ui.adsabs.harvard.edu/abs/2012yCat.2311....0C} {p. II/311}

\bibitem[\protect\citeauthoryear{{Cutri} et~al.,}{{Cutri} et~al.}{2003}]{2mass}
{Cutri} R.~M.,  et~al., 2003, VizieR Online Data Catalog, \href
  {https://ui.adsabs.harvard.edu/abs/2003yCat.2246....0C} {p. II/246}

\bibitem[\protect\citeauthoryear{{Darnley}, {Williams}, {Bode}, {Henze},
  {Ness}, {Shafter}, {Hornoch}  \& {Votruba}}{{Darnley}
  et~al.}{2014}]{Darnley14}
{Darnley} M.~J.,  {Williams} S.~C.,  {Bode} M.~F.,  {Henze} M.,  {Ness} J.~U.,
  {Shafter} A.~W.,  {Hornoch} K.,   {Votruba} V.,  2014, \mn@doi [\aap]
  {10.1051/0004-6361/201423411}, \href
  {https://ui.adsabs.harvard.edu/abs/2014A&A...563L...9D} {563, L9}

\bibitem[\protect\citeauthoryear{{Davis}, {Siess}  \& {Deschamps}}{{Davis}
  et~al.}{2013a}]{davis13}
{Davis} P.~J.,  {Siess} L.,   {Deschamps} R.,  2013a, \mn@doi [\aap]
  {10.1051/0004-6361/201220391}, \href
  {https://ui.adsabs.harvard.edu/abs/2013A&A...556A...4D} {556, A4}

\bibitem[\protect\citeauthoryear{{Davis}, {Siess}  \& {Deschamps}}{{Davis}
  et~al.}{2013b}]{Dav13}
{Davis} P.~J.,  {Siess} L.,   {Deschamps} R.,  2013b, \mn@doi [\aap]
  {10.1051/0004-6361/201220391}, \href
  {https://ui.adsabs.harvard.edu/abs/2013A&A...556A...4D} {556, A4}

\bibitem[\protect\citeauthoryear{{Dewi} \& {Tauris}}{{Dewi} \&
  {Tauris}}{2000}]{Dew00}
{Dewi} J.~D.~M.,  {Tauris} T.~M.,  2000, \aap, \href
  {http://adsabs.harvard.edu/abs/2000A%26A...360.1043D} {360, 1043}

\bibitem[\protect\citeauthoryear{{Dosopoulou} \& {Kalogera}}{{Dosopoulou} \&
  {Kalogera}}{2016a}]{Dos16}
{Dosopoulou} F.,  {Kalogera} V.,  2016a, \mn@doi [\apj]
  {10.3847/0004-637X/825/1/70}, \href
  {https://ui.adsabs.harvard.edu/abs/2016ApJ...825...70D} {825, 70}

\bibitem[\protect\citeauthoryear{{Dosopoulou} \& {Kalogera}}{{Dosopoulou} \&
  {Kalogera}}{2016b}]{Dos16b}
{Dosopoulou} F.,  {Kalogera} V.,  2016b, \mn@doi [\apj]
  {10.3847/0004-637X/825/1/71}, \href
  {https://ui.adsabs.harvard.edu/abs/2016ApJ...825...71D} {825, 71}

\bibitem[\protect\citeauthoryear{{Dotter}}{{Dotter}}{2016}]{mist0}
{Dotter} A.,  2016, \mn@doi [\apjs] {10.3847/0067-0049/222/1/8}, \href
  {https://ui.adsabs.harvard.edu/abs/2016ApJS..222....8D} {222, 8}

\bibitem[\protect\citeauthoryear{{Duch{\^e}ne} \& {Kraus}}{{Duch{\^e}ne} \&
  {Kraus}}{2013}]{Duc13}
{Duch{\^e}ne} G.,  {Kraus} A.,  2013, \mn@doi [\araa]
  {10.1146/annurev-astro-081710-102602}, \href
  {http://adsabs.harvard.edu/abs/2013ARA%26A..51..269D} {51, 269}

\bibitem[\protect\citeauthoryear{{Eggleton}}{{Eggleton}}{2009}]{Egg09}
{Eggleton} P.~P.,  2009, \mn@doi [\mnras] {10.1111/j.1365-2966.2009.15372.x},
  \href {http://adsabs.harvard.edu/abs/2009MNRAS.399.1471E} {399, 1471}

\bibitem[\protect\citeauthoryear{{El-Badry}, {Rix}  \& {Heintz}}{{El-Badry}
  et~al.}{2021}]{elbadry21}
{El-Badry} K.,  {Rix} H.-W.,   {Heintz} T.~M.,  2021, \mn@doi [\mnras]
  {10.1093/mnras/stab323}, \href
  {https://ui.adsabs.harvard.edu/abs/2021MNRAS.tmp..394E} {}

\bibitem[\protect\citeauthoryear{{Epelstain}, {Yaron}, {Kovetz}  \&
  {Prialnik}}{{Epelstain} et~al.}{2007}]{epelstain}
{Epelstain} N.,  {Yaron} O.,  {Kovetz} A.,   {Prialnik} D.,  2007, \mn@doi
  [\mnras] {10.1111/j.1365-2966.2006.11254.x}, \href
  {https://ui.adsabs.harvard.edu/abs/2007MNRAS.374.1449E} {374, 1449}

\bibitem[\protect\citeauthoryear{{Fabian}, {Pringle}  \& {Rees}}{{Fabian}
  et~al.}{1975}]{Fab75}
{Fabian} A.~C.,  {Pringle} J.~E.,   {Rees} M.~J.,  1975, \mn@doi [\mnras]
  {10.1093/mnras/172.1.15P}, \href
  {https://ui.adsabs.harvard.edu/abs/1975MNRAS.172P..15F} {172, 15}

\bibitem[\protect\citeauthoryear{{Fabrycky} \& {Tremaine}}{{Fabrycky} \&
  {Tremaine}}{2007}]{Fab07}
{Fabrycky} D.,  {Tremaine} S.,  2007, \mn@doi [\apj] {10.1086/521702}, \href
  {http://adsabs.harvard.edu/abs/2007ApJ...669.1298F} {669, 1298}

\bibitem[\protect\citeauthoryear{{Fujimoto}}{{Fujimoto}}{1982}]{fujimoto}
{Fujimoto} M.~Y.,  1982, \mn@doi [\apj] {10.1086/160030}, \href
  {https://ui.adsabs.harvard.edu/abs/1982ApJ...257..767F} {257, 767}

\bibitem[\protect\citeauthoryear{{Gaia Collaboration} et~al.,}{{Gaia
  Collaboration} et~al.}{2016}]{gaia16}
{Gaia Collaboration} et~al., 2016, \mn@doi [\aap]
  {10.1051/0004-6361/201629272}, \href
  {https://ui.adsabs.harvard.edu/abs/2016A&A...595A...1G} {595, A1}

\bibitem[\protect\citeauthoryear{{Gaia Collaboration} et~al.,}{{Gaia
  Collaboration} et~al.}{2021}]{gaia21}
{Gaia Collaboration} et~al., 2021, \mn@doi [\aap]
  {10.1051/0004-6361/202039657}, \href
  {https://ui.adsabs.harvard.edu/abs/2021A&A...649A...1G} {649, A1}

\bibitem[\protect\citeauthoryear{{Glanz} \& {Perets}}{{Glanz} \&
  {Perets}}{2018}]{Gla18}
{Glanz} H.,  {Perets} H.~B.,  2018, \mn@doi [\mnras] {10.1093/mnrasl/sly065},
  \href {https://ui.adsabs.harvard.edu/#abs/2018MNRAS.478L..12G} {478, L12}

\bibitem[\protect\citeauthoryear{{Glanz} \& {Perets}}{{Glanz} \&
  {Perets}}{2021}]{Gla21}
{Glanz} H.,  {Perets} H.~B.,  2021, arXiv e-prints, \href
  {https://ui.adsabs.harvard.edu/abs/2021arXiv210502227G} {p. arXiv:2105.02227}

\bibitem[\protect\citeauthoryear{{Gonz{\'a}lez-Fern{\'a}ndez}
  et~al.,}{{Gonz{\'a}lez-Fern{\'a}ndez} et~al.}{2018}]{vista2}
{Gonz{\'a}lez-Fern{\'a}ndez} C.,  et~al., 2018, \mn@doi [\mnras]
  {10.1093/mnras/stx3073}, \href
  {https://ui.adsabs.harvard.edu/abs/2018MNRAS.474.5459G} {474, 5459}

\bibitem[\protect\citeauthoryear{{Grishin}, {Perets}  \& {Fragione}}{{Grishin}
  et~al.}{2018}]{Gri18}
{Grishin} E.,  {Perets} H.~B.,   {Fragione} G.,  2018, \mn@doi [\mnras]
  {10.1093/mnras/sty2477}, \href
  {https://ui.adsabs.harvard.edu/abs/2018MNRAS.481.4907G} {481, 4907}

\bibitem[\protect\citeauthoryear{{Hachisu} \& {Kato}}{{Hachisu} \&
  {Kato}}{2001}]{hachisu01}
{Hachisu} I.,  {Kato} M.,  2001, \mn@doi [\apj] {10.1086/321601}, \href
  {https://ui.adsabs.harvard.edu/abs/2001ApJ...558..323H} {558, 323}

\bibitem[\protect\citeauthoryear{{Hadjidemetriou}}{{Hadjidemetriou}}{1966}]{Had66}
{Hadjidemetriou} J.,  1966, in {Kontopoulos} G.~I.,  ed., ~ Vol. 25, The Theory
  of Orbits in the Solar System and in Stellar Systems. p.~129

\bibitem[\protect\citeauthoryear{{Hamers} \& {Dosopoulou}}{{Hamers} \&
  {Dosopoulou}}{2019}]{Ham19}
{Hamers} A.~S.,  {Dosopoulou} F.,  2019, \mn@doi [\apj]
  {10.3847/1538-4357/ab001d}, \href
  {https://ui.adsabs.harvard.edu/abs/2019ApJ...872..119H} {872, 119}

\bibitem[\protect\citeauthoryear{{Hamers}, {Perets}, {Thompson}  \&
  {Neunteufel}}{{Hamers} et~al.}{2021}]{Ham21}
{Hamers} A.~S.,  {Perets} H.~B.,  {Thompson} T.~A.,   {Neunteufel} P.,  2021,
  arXiv e-prints, \href {https://ui.adsabs.harvard.edu/abs/2021arXiv210713620H}
  {p. arXiv:2107.13620}

\bibitem[\protect\citeauthoryear{{He} \& {Petrovich}}{{He} \&
  {Petrovich}}{2018}]{He18}
{He} M.~Y.,  {Petrovich} C.,  2018, \mn@doi [\mnras] {10.1093/mnras/stx2718},
  \href {https://ui.adsabs.harvard.edu/abs/2018MNRAS.474...20H} {474, 20}

\bibitem[\protect\citeauthoryear{{Heggie}}{{Heggie}}{1975}]{Heg75}
{Heggie} D.~C.,  1975, \mnras, \href
  {http://adsabs.harvard.edu/abs/1975MNRAS.173..729H} {173, 729}

\bibitem[\protect\citeauthoryear{{Henden}, {Levine}, {Terrell}, {Smith}  \&
  {Welch}}{{Henden} et~al.}{2012}]{apass1}
{Henden} A.~A.,  {Levine} S.~E.,  {Terrell} D.,  {Smith} T.~C.,   {Welch} D.,
  2012, \jaavso, \href {https://ui.adsabs.harvard.edu/abs/2012JAVSO..40..430H}
  {40, 430}

\bibitem[\protect\citeauthoryear{{Hurley}, {Pols}  \& {Tout}}{{Hurley}
  et~al.}{2000}]{Hur00}
{Hurley} J.~R.,  {Pols} O.~R.,   {Tout} C.~A.,  2000, \mn@doi [\mnras]
  {10.1046/j.1365-8711.2000.03426.x}, \href
  {https://ui.adsabs.harvard.edu/abs/2000MNRAS.315..543H} {315, 543}

\bibitem[\protect\citeauthoryear{{Hurley}, {Tout}  \& {Pols}}{{Hurley}
  et~al.}{2002}]{Hur02}
{Hurley} J.~R.,  {Tout} C.~A.,   {Pols} O.~R.,  2002, \mn@doi [\mnras]
  {10.1046/j.1365-8711.2002.05038.x}, \href
  {https://ui.adsabs.harvard.edu/abs/2002MNRAS.329..897H} {329, 897}

\bibitem[\protect\citeauthoryear{{Hut}}{{Hut}}{1981}]{Hut81}
{Hut} P.,  1981, \aap, \href
  {https://ui.adsabs.harvard.edu/abs/1981A&A....99..126H} {99, 126}

\bibitem[\protect\citeauthoryear{{Iben} \& {Tutukov}}{{Iben} \&
  {Tutukov}}{1999}]{Ibe99}
{Iben} Icko J.,  {Tutukov} A.~V.,  1999, \mn@doi [\apj] {10.1086/306672}, \href
  {https://ui.adsabs.harvard.edu/abs/1999ApJ...511..324I} {511, 324}

\bibitem[\protect\citeauthoryear{{Igoshev}, {Perets}  \& {Michaely}}{{Igoshev}
  et~al.}{2020}]{Igo20}
{Igoshev} A.~P.,  {Perets} H.~B.,   {Michaely} E.,  2020, \mn@doi [\mnras]
  {10.1093/mnras/staa833}, \href
  {https://ui.adsabs.harvard.edu/abs/2020MNRAS.494.1448I} {494, 1448}

\bibitem[\protect\citeauthoryear{{Ivanova} et~al.,}{{Ivanova}
  et~al.}{2013}]{Iva13}
{Ivanova} N.,  et~al., 2013, \mn@doi [\aapr] {10.1007/s00159-013-0059-2}, \href
  {https://ui.adsabs.harvard.edu/abs/2013A&ARv..21...59I} {21, 59}

\bibitem[\protect\citeauthoryear{{Kato}, {Saio}, {Hachisu}  \& {Nomoto}}{{Kato}
  et~al.}{2014}]{kato-nova}
{Kato} M.,  {Saio} H.,  {Hachisu} I.,   {Nomoto} K.,  2014, \mn@doi [\apj]
  {10.1088/0004-637X/793/2/136}, \href
  {https://ui.adsabs.harvard.edu/abs/2014ApJ...793..136K} {793, 136}

\bibitem[\protect\citeauthoryear{{Katz} \& {Dong}}{{Katz} \&
  {Dong}}{2012}]{Kat12}
{Katz} B.,  {Dong} S.,  2012, preprint, \href
  {http://adsabs.harvard.edu/abs/2012arXiv1211.4584K} {} (\mn@eprint {arXiv}
  {1211.4584})

\bibitem[\protect\citeauthoryear{{King} \& {van Teeseling}}{{King} \& {van
  Teeseling}}{1998}]{king98}
{King} A.~R.,  {van Teeseling} A.,  1998, \aap, \href
  {https://ui.adsabs.harvard.edu/abs/1998A&A...338..965K} {338, 965}

\bibitem[\protect\citeauthoryear{{Kinoshita} \& {Nakai}}{{Kinoshita} \&
  {Nakai}}{1999}]{Kin99}
{Kinoshita} H.,  {Nakai} H.,  1999, \mn@doi [Celestial Mechanics and Dynamical
  Astronomy] {10.1023/A:1008321310187}, \href
  {https://ui.adsabs.harvard.edu/abs/1999CeMDA..75..125K} {75, 125}

\bibitem[\protect\citeauthoryear{{Kiseleva}, {Eggleton}  \& {Orlov}}{{Kiseleva}
  et~al.}{1994}]{Kis94}
{Kiseleva} L.~G.,  {Eggleton} P.~P.,   {Orlov} V.~V.,  1994, \mn@doi [\mnras]
  {10.1093/mnras/270.4.936}, \href
  {https://ui.adsabs.harvard.edu/abs/1994MNRAS.270..936K} {270, 936}

\bibitem[\protect\citeauthoryear{{Kiseleva}, {Eggleton}  \&
  {Mikkola}}{{Kiseleva} et~al.}{1998}]{Kis98}
{Kiseleva} L.~G.,  {Eggleton} P.~P.,   {Mikkola} S.,  1998, \mn@doi [\mnras]
  {10.1046/j.1365-8711.1998.01903.x}, \href
  {http://adsabs.harvard.edu/abs/1998MNRAS.300..292K} {300, 292}

\bibitem[\protect\citeauthoryear{{Knigge}}{{Knigge}}{2006}]{knigge06}
{Knigge} C.,  2006, \mn@doi [\mnras] {10.1111/j.1365-2966.2006.11096.x}, \href
  {https://ui.adsabs.harvard.edu/abs/2006MNRAS.373..484K} {373, 484}

\bibitem[\protect\citeauthoryear{{Knigge}, {King}  \& {Patterson}}{{Knigge}
  et~al.}{2000}]{Knigge00}
{Knigge} C.,  {King} A.~R.,   {Patterson} J.,  2000, \aap, \href
  {https://ui.adsabs.harvard.edu/abs/2000A&A...364L..75K} {364, L75}

\bibitem[\protect\citeauthoryear{{Knigge}, {Baraffe}  \& {Patterson}}{{Knigge}
  et~al.}{2011}]{knigge11}
{Knigge} C.,  {Baraffe} I.,   {Patterson} J.,  2011, \mn@doi [\apjs]
  {10.1088/0067-0049/194/2/28}, \href
  {https://ui.adsabs.harvard.edu/abs/2011ApJS..194...28K} {194, 28}

\bibitem[\protect\citeauthoryear{{Kochanek}}{{Kochanek}}{1992}]{Koc92}
{Kochanek} C.~S.,  1992, \mn@doi [\apj] {10.1086/170966}, \href
  {https://ui.adsabs.harvard.edu/abs/1992ApJ...385..604K} {385, 604}

\bibitem[\protect\citeauthoryear{{Kozai}}{{Kozai}}{1962}]{Koz62}
{Kozai} Y.,  1962, \mn@doi [\aj] {10.1086/108790}, \href
  {http://adsabs.harvard.edu/abs/1962AJ.....67..591K} {67, 591}

\bibitem[\protect\citeauthoryear{{Kroupa}, {Tout}  \& {Gilmore}}{{Kroupa}
  et~al.}{1993}]{Kro93}
{Kroupa} P.,  {Tout} C.~A.,   {Gilmore} G.,  1993, \mnras, \href
  {http://adsabs.harvard.edu/abs/1993MNRAS.262..545K} {262, 545}

\bibitem[\protect\citeauthoryear{{Lajoie} \& {Sills}}{{Lajoie} \&
  {Sills}}{2011a}]{lajoie}
{Lajoie} C.-P.,  {Sills} A.,  2011a, \mn@doi [\apj]
  {10.1088/0004-637X/726/2/67}, \href
  {https://ui.adsabs.harvard.edu/abs/2011ApJ...726...67L} {726, 67}

\bibitem[\protect\citeauthoryear{{Lajoie} \& {Sills}}{{Lajoie} \&
  {Sills}}{2011b}]{Laj11}
{Lajoie} C.-P.,  {Sills} A.,  2011b, \mn@doi [\apj]
  {10.1088/0004-637X/726/2/67}, \href
  {https://ui.adsabs.harvard.edu/abs/2011ApJ...726...67L} {726, 67}

\bibitem[\protect\citeauthoryear{{Law-Smith} et~al.,}{{Law-Smith}
  et~al.}{2020}]{Law20}
{Law-Smith} J. A.~P.,  et~al., 2020, arXiv e-prints, \href
  {https://ui.adsabs.harvard.edu/abs/2020arXiv201106630L} {p. arXiv:2011.06630}

\bibitem[\protect\citeauthoryear{{Layton}, {Blondin}, {Owen}  \&
  {Stevens}}{{Layton} et~al.}{1998}]{Lay98}
{Layton} J.~T.,  {Blondin} J.~M.,  {Owen} M.~P.,   {Stevens} I.~R.,  1998,
  \mn@doi [\na] {10.1016/S1384-1076(97)00047-X}, \href
  {https://ui.adsabs.harvard.edu/abs/1998NewA....3..111L} {3, 111}

\bibitem[\protect\citeauthoryear{{Levine}}{{Levine}}{2017}]{apass2}
{Levine} S.,  2017, \jaavso, \href
  {https://ui.adsabs.harvard.edu/abs/2017JAVSO..45..127L} {45, 127}

\bibitem[\protect\citeauthoryear{{Lidov}}{{Lidov}}{1962}]{Lid62}
{Lidov} M.~L.,  1962, \mn@doi [\planss] {10.1016/0032-0633(62)90129-0}, \href
  {http://adsabs.harvard.edu/abs/1962P%26SS....9..719L} {9, 719}

\bibitem[\protect\citeauthoryear{{Liu} \& {Lai}}{{Liu} \& {Lai}}{2018}]{Liu18}
{Liu} B.,  {Lai} D.,  2018, \mn@doi [\apj] {10.3847/1538-4357/aad09f}, \href
  {https://ui.adsabs.harvard.edu/abs/2018ApJ...863...68L} {863, 68}

\bibitem[\protect\citeauthoryear{{Liu}, {Mu{\~n}oz}  \& {Lai}}{{Liu}
  et~al.}{2015}]{Liu15}
{Liu} B.,  {Mu{\~n}oz} D.~J.,   {Lai} D.,  2015, \mn@doi [\mnras]
  {10.1093/mnras/stu2396}, \href
  {http://adsabs.harvard.edu/abs/2015MNRAS.447..747L} {447, 747}

\bibitem[\protect\citeauthoryear{{Livio} \& {Mazzali}}{{Livio} \&
  {Mazzali}}{2018}]{livio18}
{Livio} M.,  {Mazzali} P.,  2018, \mn@doi [Phys. Rep.]
  {10.1016/j.physrep.2018.02.002}, \href
  {https://ui.adsabs.harvard.edu/abs/2018PhR...736....1L} {736, 1}

\bibitem[\protect\citeauthoryear{{Livio} \& {Soker}}{{Livio} \&
  {Soker}}{1988}]{Liv88}
{Livio} M.,  {Soker} N.,  1988, \mn@doi [\apj] {10.1086/166419}, \href
  {http://adsabs.harvard.edu/abs/1988ApJ...329..764L} {329, 764}

\bibitem[\protect\citeauthoryear{{Livio} \& {Truran}}{{Livio} \&
  {Truran}}{1992}]{livio92}
{Livio} M.,  {Truran} J.~W.,  1992, \mn@doi [\apj] {10.1086/171242}, \href
  {https://ui.adsabs.harvard.edu/abs/1992ApJ...389..695L} {389, 695}

\bibitem[\protect\citeauthoryear{{Loveridge}, {van der Sluys}  \&
  {Kalogera}}{{Loveridge} et~al.}{2011}]{Lov11}
{Loveridge} A.~J.,  {van der Sluys} M.~V.,   {Kalogera} V.,  2011, \mn@doi
  [\apj] {10.1088/0004-637X/743/1/49}, \href
  {http://adsabs.harvard.edu/abs/2011ApJ...743...49L} {743, 49}

\bibitem[\protect\citeauthoryear{{Luo}, {Katz}  \& {Dong}}{{Luo}
  et~al.}{2016}]{Luo16}
{Luo} L.,  {Katz} B.,   {Dong} S.,  2016, \mn@doi [\mnras]
  {10.1093/mnras/stw475}, \href
  {https://ui.adsabs.harvard.edu/abs/2016MNRAS.458.3060L} {458, 3060}

\bibitem[\protect\citeauthoryear{{Mardling}}{{Mardling}}{1995}]{Mar95}
{Mardling} R.~A.,  1995, \mn@doi [\apj] {10.1086/176179}, \href
  {https://ui.adsabs.harvard.edu/abs/1995ApJ...450..732M} {450, 732}

\bibitem[\protect\citeauthoryear{{Mardling} \& {Aarseth}}{{Mardling} \&
  {Aarseth}}{1999}]{Mar99}
{Mardling} R.,  {Aarseth} S.,  1999, in {Steves} B.~A.,  {Roy} A.~E.,  eds,
  NATO Advanced Science Institutes (ASI) Series C Vol. 522, NATO Advanced
  Science Institutes (ASI) Series C. p.~385

\bibitem[\protect\citeauthoryear{{Mazeh} \& {Shaham}}{{Mazeh} \&
  {Shaham}}{1979}]{Maz79}
{Mazeh} T.,  {Shaham} J.,  1979, \aap, \href
  {http://adsabs.harvard.edu/abs/1979A%26A....77..145M} {77, 145}

\bibitem[\protect\citeauthoryear{{McMahon}, {Banerji}, {Gonzalez}, {Koposov},
  {Bejar}, {Lodieu}, {Rebolo}  \& {VHS Collaboration}}{{McMahon}
  et~al.}{2013}]{vista1}
{McMahon} R.~G.,  {Banerji} M.,  {Gonzalez} E.,  {Koposov} S.~E.,  {Bejar}
  V.~J.,  {Lodieu} N.,  {Rebolo} R.,   {VHS Collaboration} 2013, The Messenger,
  \href {https://ui.adsabs.harvard.edu/abs/2013Msngr.154...35M} {154, 35}

\bibitem[\protect\citeauthoryear{{Michaely} \& {Perets}}{{Michaely} \&
  {Perets}}{2014}]{Michaely14}
{Michaely} E.,  {Perets} H.~B.,  2014, \mn@doi [\apj]
  {10.1088/0004-637X/794/2/122}, \href
  {https://ui.adsabs.harvard.edu/abs/2014ApJ...794..122M} {794, 122}

\bibitem[\protect\citeauthoryear{{Michaely} \& {Perets}}{{Michaely} \&
  {Perets}}{2019}]{Mic19}
{Michaely} E.,  {Perets} H.~B.,  2019, \mn@doi [\mnras] {10.1093/mnras/stz352},
  \href {https://ui.adsabs.harvard.edu/abs/2019MNRAS.484.4711M} {484, 4711}

\bibitem[\protect\citeauthoryear{{Moe} \& {Di Stefano}}{{Moe} \& {Di
  Stefano}}{2017}]{Moe17}
{Moe} M.,  {Di Stefano} R.,  2017, \mn@doi [\apjs] {10.3847/1538-4365/aa6fb6},
  \href {http://adsabs.harvard.edu/abs/2017ApJS..230...15M} {230, 15}

\bibitem[\protect\citeauthoryear{{Moe} \& {Kratter}}{{Moe} \&
  {Kratter}}{2018}]{Moe18}
{Moe} M.,  {Kratter} K.~M.,  2018, \mn@doi [\apj] {10.3847/1538-4357/aaa6d2},
  \href {https://ui.adsabs.harvard.edu/abs/2018ApJ...854...44M} {854, 44}

\bibitem[\protect\citeauthoryear{{Nandez} \& {Ivanova}}{{Nandez} \&
  {Ivanova}}{2016}]{Nan16}
{Nandez} J.~L.~A.,  {Ivanova} N.,  2016, \mn@doi [\mnras]
  {10.1093/mnras/stw1266}, \href
  {https://ui.adsabs.harvard.edu/abs/2016MNRAS.460.3992N} {460, 3992}

\bibitem[\protect\citeauthoryear{{Naoz}}{{Naoz}}{2016}]{Nao16}
{Naoz} S.,  2016, \mn@doi [\araa] {10.1146/annurev-astro-081915-023315}, \href
  {https://ui.adsabs.harvard.edu/abs/2016ARA&A..54..441N} {54, 441}

\bibitem[\protect\citeauthoryear{{Nelemans}, {Verbunt}, {Yungelson}  \&
  {Portegies Zwart}}{{Nelemans} et~al.}{2000}]{Nel00}
{Nelemans} G.,  {Verbunt} F.,  {Yungelson} L.~R.,   {Portegies Zwart} S.~F.,
  2000, \aap, \href {http://adsabs.harvard.edu/abs/2000A%26A...360.1011N} {360,
  1011}

\bibitem[\protect\citeauthoryear{{Nelemans}, {Siess}, {Repetto}, {Toonen}  \&
  {Phinney}}{{Nelemans} et~al.}{2016}]{nelemans16}
{Nelemans} G.,  {Siess} L.,  {Repetto} S.,  {Toonen} S.,   {Phinney} E.~S.,
  2016, \mn@doi [\apj] {10.3847/0004-637X/817/1/69}, \href
  {https://ui.adsabs.harvard.edu/abs/2016ApJ...817...69N} {817, 69}

\bibitem[\protect\citeauthoryear{{Nomoto}}{{Nomoto}}{1982}]{nomoto}
{Nomoto} K.,  1982, \mn@doi [\apj] {10.1086/159682}, \href
  {https://ui.adsabs.harvard.edu/abs/1982ApJ...253..798N} {253, 798}

\bibitem[\protect\citeauthoryear{{Nomoto} \& {et al.}}{{Nomoto} \& {et
  al.}}{2000}]{nomoto00}
{Nomoto} K.,  {et al.} 2000, in {Niemeyer} J.~C.,  {Truran} J.~W.,  eds, Type
  Ia Supernovae, Theory and Cosmology. p.~63 (\mn@eprint {arXiv}
  {astro-ph/9907386})

\bibitem[\protect\citeauthoryear{{Paczynski}}{{Paczynski}}{1976}]{Pac76}
{Paczynski} B.,  1976, in {P.~Eggleton, S.~Mitton, \& J.~Whelan} ed.,  IAU
  Symposium Vol. 73, Structure and Evolution of Close Binary Systems. Kluwer,
  Dordrecht, p.~75

\bibitem[\protect\citeauthoryear{{Page} et~al.,}{{Page} et~al.}{2012}]{xmmom}
{Page} M.~J.,  et~al., 2012, \mn@doi [\mnras]
  {10.1111/j.1365-2966.2012.21706.x}, \href
  {https://ui.adsabs.harvard.edu/abs/2012MNRAS.426..903P} {426, 903}

\bibitem[\protect\citeauthoryear{{Pagnotta} \& {Schaefer}}{{Pagnotta} \&
  {Schaefer}}{2014}]{Pagnotta14}
{Pagnotta} A.,  {Schaefer} B.~E.,  2014, \mn@doi [\apj]
  {10.1088/0004-637X/788/2/164}, \href
  {https://ui.adsabs.harvard.edu/abs/2014ApJ...788..164P} {788, 164}

\bibitem[\protect\citeauthoryear{{Pala} et~al.,}{{Pala} et~al.}{2020}]{pala20}
{Pala} A.~F.,  et~al., 2020, \mn@doi [\mnras] {10.1093/mnras/staa764}, \href
  {https://ui.adsabs.harvard.edu/abs/2020MNRAS.494.3799P} {494, 3799}

\bibitem[\protect\citeauthoryear{{Patterson} et~al.,}{{Patterson}
  et~al.}{1998}]{patterson98}
{Patterson} J.,  et~al., 1998, \mn@doi [\pasp] {10.1086/316147}, \href
  {https://ui.adsabs.harvard.edu/abs/1998PASP..110..380P} {110, 380}

\bibitem[\protect\citeauthoryear{{Patterson} et~al.,}{{Patterson}
  et~al.}{2017}]{Patterson17}
{Patterson} J.,  et~al., 2017, \mn@doi [\mnras] {10.1093/mnras/stw2970}, \href
  {https://ui.adsabs.harvard.edu/abs/2017MNRAS.466..581P} {466, 581}

\bibitem[\protect\citeauthoryear{{Patterson} et~al.,}{{Patterson}
  et~al.}{2020}]{Patterson20}
{Patterson} J.,  et~al., 2020, arXiv e-prints, \href
  {https://ui.adsabs.harvard.edu/abs/2020arXiv201007812P} {p. arXiv:2010.07812}

\bibitem[\protect\citeauthoryear{{Perets} \& {Kratter}}{{Perets} \&
  {Kratter}}{2012}]{Perets12}
{Perets} H.~B.,  {Kratter} K.~M.,  2012, \mn@doi [\apj]
  {10.1088/0004-637X/760/2/99}, \href
  {https://ui.adsabs.harvard.edu/abs/2012ApJ...760...99P} {760, 99}

\bibitem[\protect\citeauthoryear{{Portegies Zwart} \& {McMillan}}{{Portegies
  Zwart} \& {McMillan}}{2018}]{Por18}
{Portegies Zwart} S.,  {McMillan} S.,  2018, {Astrophysical Recipes; The art of
  AMUSE}, \mn@doi{10.1088/978-0-7503-1320-9.
}

\bibitem[\protect\citeauthoryear{{Press} \& {Teukolsky}}{{Press} \&
  {Teukolsky}}{1977}]{Pre77}
{Press} W.~H.,  {Teukolsky} S.~A.,  1977, \mn@doi [\apj] {10.1086/155143},
  \href {https://ui.adsabs.harvard.edu/abs/1977ApJ...213..183P} {213, 183}

\bibitem[\protect\citeauthoryear{{Raghavan} et~al.,}{{Raghavan}
  et~al.}{2010}]{Rag10}
{Raghavan} D.,  et~al., 2010, \mn@doi [\apjs] {10.1088/0067-0049/190/1/1},
  \href {http://adsabs.harvard.edu/abs/2010ApJS..190....1R} {190, 1}

\bibitem[\protect\citeauthoryear{{Rappaport}, {Verbunt}  \& {Joss}}{{Rappaport}
  et~al.}{1983}]{Rap83}
{Rappaport} S.,  {Verbunt} F.,   {Joss} P.~C.,  1983, \mn@doi [\apj]
  {10.1086/161569}, \href
  {https://ui.adsabs.harvard.edu/abs/1983ApJ...275..713R} {275, 713}

\bibitem[\protect\citeauthoryear{{Regev} \& {Shara}}{{Regev} \&
  {Shara}}{1987}]{Reg87}
{Regev} O.,  {Shara} M.~M.,  1987, \mn@doi [\mnras] {10.1093/mnras/227.4.967},
  \href {https://ui.adsabs.harvard.edu/abs/1987MNRAS.227..967R} {227, 967}

\bibitem[\protect\citeauthoryear{{Reg{\"o}s}, {Bailey}  \&
  {Mardling}}{{Reg{\"o}s} et~al.}{2005}]{Reg05}
{Reg{\"o}s} E.,  {Bailey} V.~C.,   {Mardling} R.,  2005, \mn@doi [\mnras]
  {10.1111/j.1365-2966.2005.08813.x}, \href
  {https://ui.adsabs.harvard.edu/abs/2005MNRAS.358..544R} {358, 544}

\bibitem[\protect\citeauthoryear{{Reichardt}, {De Marco}, {Iaconi}, {Chamandy}
  \& {Price}}{{Reichardt} et~al.}{2020}]{Rei20}
{Reichardt} T.~A.,  {De Marco} O.,  {Iaconi} R.,  {Chamandy} L.,   {Price}
  D.~J.,  2020, \mn@doi [\mnras] {10.1093/mnras/staa937}, \href
  {https://ui.adsabs.harvard.edu/abs/2020MNRAS.494.5333R} {494, 5333}

\bibitem[\protect\citeauthoryear{{Remage Evans}}{{Remage Evans}}{2011}]{Rem11}
{Remage Evans} N.,  2011, Bulletin de la Societe Royale des Sciences de Liege,
  \href {http://adsabs.harvard.edu/abs/2011BSRSL..80..663E} {80, 663}

\bibitem[\protect\citeauthoryear{{Riello} et~al.,}{{Riello}
  et~al.}{2021}]{gaiaphotom}
{Riello} M.,  et~al., 2021, \mn@doi [\aap] {10.1051/0004-6361/202039587}, \href
  {https://ui.adsabs.harvard.edu/abs/2021A&A...649A...3R} {649, A3}

\bibitem[\protect\citeauthoryear{{Rodriguez} \& {Antonini}}{{Rodriguez} \&
  {Antonini}}{2018}]{Rod18}
{Rodriguez} C.~L.,  {Antonini} F.,  2018, \mn@doi [\apj]
  {10.3847/1538-4357/aacea4}, \href
  {https://ui.adsabs.harvard.edu/abs/2018ApJ...863....7R} {863, 7}

\bibitem[\protect\citeauthoryear{{Rozyczka}, {Yorke}, {Bodenheimer}, {Mueller}
  \& {Hashimoto}}{{Rozyczka} et~al.}{1989}]{Roz89}
{Rozyczka} M.,  {Yorke} H.~W.,  {Bodenheimer} P.,  {Mueller} E.,   {Hashimoto}
  M.,  1989, \aap, \href
  {https://ui.adsabs.harvard.edu/abs/1989A&A...208...69R} {208, 69}

\bibitem[\protect\citeauthoryear{{Sana} et~al.,}{{Sana} et~al.}{2012}]{San12}
{Sana} H.,  et~al., 2012, \mn@doi [Science] {10.1126/science.1223344}, \href
  {http://adsabs.harvard.edu/abs/2012Sci...337..444S} {337, 444}

\bibitem[\protect\citeauthoryear{{Schaefer}}{{Schaefer}}{2010}]{Schaefer10}
{Schaefer} B.~E.,  2010, \mn@doi [\apjs] {10.1088/0067-0049/187/2/275}, \href
  {https://ui.adsabs.harvard.edu/abs/2010ApJS..187..275S} {187, 275}

\bibitem[\protect\citeauthoryear{{Schaefer}, {Pagnotta}  \& {Shara}}{{Schaefer}
  et~al.}{2010}]{schaefer10a}
{Schaefer} B.~E.,  {Pagnotta} A.,   {Shara} M.~M.,  2010, \mn@doi [\apj]
  {10.1088/0004-637X/708/1/381}, \href
  {https://ui.adsabs.harvard.edu/abs/2010ApJ...708..381S} {708, 381}

\bibitem[\protect\citeauthoryear{{Sepinsky}, {Willems}, {Kalogera}  \&
  {Rasio}}{{Sepinsky} et~al.}{2007}]{Sep07}
{Sepinsky} J.~F.,  {Willems} B.,  {Kalogera} V.,   {Rasio} F.~A.,  2007,
  \mn@doi [\apj] {10.1086/520911}, \href
  {https://ui.adsabs.harvard.edu/abs/2007ApJ...667.1170S} {667, 1170}

\bibitem[\protect\citeauthoryear{{Sepinsky}, {Willems}, {Kalogera}  \&
  {Rasio}}{{Sepinsky} et~al.}{2009}]{Sep09}
{Sepinsky} J.~F.,  {Willems} B.,  {Kalogera} V.,   {Rasio} F.~A.,  2009,
  \mn@doi [\apj] {10.1088/0004-637X/702/2/1387}, \href
  {https://ui.adsabs.harvard.edu/abs/2009ApJ...702.1387S} {702, 1387}

\bibitem[\protect\citeauthoryear{{Sepinsky}, {Willems}, {Kalogera}  \&
  {Rasio}}{{Sepinsky} et~al.}{2010}]{Sep10}
{Sepinsky} J.~F.,  {Willems} B.,  {Kalogera} V.,   {Rasio} F.~A.,  2010,
  \mn@doi [\apj] {10.1088/0004-637X/724/1/546}, \href
  {https://ui.adsabs.harvard.edu/abs/2010ApJ...724..546S} {724, 546}

\bibitem[\protect\citeauthoryear{{Shappee} \& {Thompson}}{{Shappee} \&
  {Thompson}}{2013}]{Shappee13}
{Shappee} B.~J.,  {Thompson} T.~A.,  2013, \mn@doi [\apj]
  {10.1088/0004-637X/766/1/64}, \href
  {https://ui.adsabs.harvard.edu/abs/2013ApJ...766...64S} {766, 64}

\bibitem[\protect\citeauthoryear{{Shara} \& {Regev}}{{Shara} \&
  {Regev}}{1986}]{Sha86}
{Shara} M.~M.,  {Regev} O.,  1986, \mn@doi [\apj] {10.1086/164364}, \href
  {https://ui.adsabs.harvard.edu/abs/1986ApJ...306..543S} {306, 543}

\bibitem[\protect\citeauthoryear{{Shiber}, {Iaconi}, {De Marco}  \&
  {Soker}}{{Shiber} et~al.}{2019}]{Shi19}
{Shiber} S.,  {Iaconi} R.,  {De Marco} O.,   {Soker} N.,  2019, \mn@doi
  [\mnras] {10.1093/mnras/stz2013}, \href
  {https://ui.adsabs.harvard.edu/abs/2019MNRAS.488.5615S} {488, 5615}

\bibitem[\protect\citeauthoryear{{Soker}}{{Soker}}{2004}]{Sok04}
{Soker} N.,  2004, \mn@doi [\na] {10.1016/j.newast.2004.01.004}, \href
  {http://adsabs.harvard.edu/abs/2004NewA....9..399S} {9, 399}

\bibitem[\protect\citeauthoryear{{Soker}}{{Soker}}{2015}]{Sok15}
{Soker} N.,  2015, \mn@doi [\apj] {10.1088/0004-637X/800/2/114}, \href
  {http://adsabs.harvard.edu/abs/2015ApJ...800..114S} {800, 114}

\bibitem[\protect\citeauthoryear{{Soker}, {Regev}, {Livio}  \& {Shara}}{{Soker}
  et~al.}{1987}]{Sok87}
{Soker} N.,  {Regev} O.,  {Livio} M.,   {Shara} M.~M.,  1987, \mn@doi [\apj]
  {10.1086/165409}, \href
  {https://ui.adsabs.harvard.edu/abs/1987ApJ...318..760S} {318, 760}

\bibitem[\protect\citeauthoryear{{Tokovinin}}{{Tokovinin}}{2014}]{Tok14b}
{Tokovinin} A.,  2014, \mn@doi [\aj] {10.1088/0004-6256/147/4/87}, \href
  {http://adsabs.harvard.edu/abs/2014AJ....147...87T} {147, 87}

\bibitem[\protect\citeauthoryear{{Tokovinin}}{{Tokovinin}}{2017}]{Tok17}
{Tokovinin} A.,  2017, \mn@doi [\apj] {10.3847/1538-4357/aa7746}, \href
  {http://adsabs.harvard.edu/abs/2017ApJ...844..103T} {844, 103}

\bibitem[\protect\citeauthoryear{{Tokovinin}}{{Tokovinin}}{2018}]{tokovinin18}
{Tokovinin} A.,  2018, \mn@doi [\apjs] {10.3847/1538-4365/aaa1a5}, \href
  {https://ui.adsabs.harvard.edu/abs/2018ApJS..235....6T} {235, 6}

\bibitem[\protect\citeauthoryear{{Toonen} \& {Nelemans}}{{Toonen} \&
  {Nelemans}}{2013}]{Too13}
{Toonen} S.,  {Nelemans} G.,  2013, \mn@doi [\aap]
  {10.1051/0004-6361/201321753}, \href
  {http://adsabs.harvard.edu/abs/2013A%26A...557A..87T} {557, A87}

\bibitem[\protect\citeauthoryear{{Toonen}, {Nelemans}  \& {Portegies
  Zwart}}{{Toonen} et~al.}{2012}]{Too12}
{Toonen} S.,  {Nelemans} G.,   {Portegies Zwart} S.,  2012, \mn@doi [\aap]
  {10.1051/0004-6361/201218966}, \href
  {https://ui.adsabs.harvard.edu/abs/2012A&A...546A..70T} {546, A70}

\bibitem[\protect\citeauthoryear{{Toonen}, {Voss}  \& {Knigge}}{{Toonen}
  et~al.}{2014a}]{Too14b}
{Toonen} S.,  {Voss} R.,   {Knigge} C.,  2014a, \mn@doi [\mnras]
  {10.1093/mnras/stu569}, \href
  {https://ui.adsabs.harvard.edu/abs/2014MNRAS.441..354T} {441, 354}

\bibitem[\protect\citeauthoryear{{Toonen}, {Claeys}, {Mennekens}  \&
  {Ruiter}}{{Toonen} et~al.}{2014b}]{Too14}
{Toonen} S.,  {Claeys} J.~S.~W.,  {Mennekens} N.,   {Ruiter} A.~J.,  2014b,
  \mn@doi [\aap] {10.1051/0004-6361/201321576}, \href
  {http://adsabs.harvard.edu/abs/2014A%26A...562A..14T} {562, A14}

\bibitem[\protect\citeauthoryear{{Toonen}, {Hamers}  \& {Portegies
  Zwart}}{{Toonen} et~al.}{2016}]{Toonen16}
{Toonen} S.,  {Hamers} A.,   {Portegies Zwart} S.,  2016, \mn@doi
  [Computational Astrophysics and Cosmology] {10.1186/s40668-016-0019-0}, \href
  {https://ui.adsabs.harvard.edu/abs/2016ComAC...3....6T} {3, 6}

\bibitem[\protect\citeauthoryear{{Toonen}, {Hollands}, {G{\"a}nsicke}  \&
  {Boekholt}}{{Toonen} et~al.}{2017}]{Too17}
{Toonen} S.,  {Hollands} M.,  {G{\"a}nsicke} B.~T.,   {Boekholt} T.,  2017,
  \mn@doi [\aap] {10.1051/0004-6361/201629978}, \href
  {http://adsabs.harvard.edu/abs/2017A%26A...602A..16T} {602, A16}

\bibitem[\protect\citeauthoryear{{Toonen}, {Portegies Zwart}, {Hamers}  \&
  {Bandopadhyay}}{{Toonen} et~al.}{2020}]{Too20}
{Toonen} S.,  {Portegies Zwart} S.,  {Hamers} A.~S.,   {Bandopadhyay} D.,
  2020, \mn@doi [\aap] {10.1051/0004-6361/201936835}, \href
  {https://ui.adsabs.harvard.edu/abs/2020A&A...640A..16T} {640, A16}

\bibitem[\protect\citeauthoryear{{Toonen}, {Boekholt}  \& {Portegies
  Zwart}}{{Toonen} et~al.}{2021}]{Too21}
{Toonen} S.,  {Boekholt} T.~C.~N.,   {Portegies Zwart} S.,  2021, arXiv
  e-prints, \href {https://ui.adsabs.harvard.edu/abs/2021arXiv210804272T} {p.
  arXiv:2108.04272}

\bibitem[\protect\citeauthoryear{{Townsley} \& {Bildsten}}{{Townsley} \&
  {Bildsten}}{2005}]{townsley05}
{Townsley} D.~M.,  {Bildsten} L.,  2005, \mn@doi [\apj] {10.1086/430594}, \href
  {https://ui.adsabs.harvard.edu/abs/2005ApJ...628..395T} {628, 395}

\bibitem[\protect\citeauthoryear{{Uthas}, {Knigge}  \& {Steeghs}}{{Uthas}
  et~al.}{2010}]{uthas10}
{Uthas} H.,  {Knigge} C.,   {Steeghs} D.,  2010, \mn@doi [\mnras]
  {10.1111/j.1365-2966.2010.17046.x}, \href
  {https://ui.adsabs.harvard.edu/abs/2010MNRAS.409..237U} {409, 237}

\bibitem[\protect\citeauthoryear{{Veras}, {Wyatt}, {Mustill}, {Bonsor}  \&
  {Eldridge}}{{Veras} et~al.}{2011}]{Ver11}
{Veras} D.,  {Wyatt} M.~C.,  {Mustill} A.~J.,  {Bonsor} A.,   {Eldridge} J.~J.,
   2011, \mn@doi [\mnras] {10.1111/j.1365-2966.2011.19393.x}, \href
  {https://ui.adsabs.harvard.edu/abs/2011MNRAS.417.2104V} {417, 2104}

\bibitem[\protect\citeauthoryear{{Webbink}}{{Webbink}}{1984}]{Web84}
{Webbink} R.~F.,  1984, \mn@doi [\apj] {10.1086/161701}, \href
  {http://adsabs.harvard.edu/abs/1984ApJ...277..355W} {277, 355}

\bibitem[\protect\citeauthoryear{{Webbink}, {Livio}, {Truran}  \&
  {Orio}}{{Webbink} et~al.}{1987}]{webbink87}
{Webbink} R.~F.,  {Livio} M.,  {Truran} J.~W.,   {Orio} M.,  1987, \mn@doi
  [\apj] {10.1086/165095}, \href
  {https://ui.adsabs.harvard.edu/abs/1987ApJ...314..653W} {314, 653}

\bibitem[\protect\citeauthoryear{{Wolf}, {Bildsten}, {Brooks}  \&
  {Paxton}}{{Wolf} et~al.}{2013}]{wolf}
{Wolf} W.~M.,  {Bildsten} L.,  {Brooks} J.,   {Paxton} B.,  2013, \mn@doi
  [\apj] {10.1088/0004-637X/777/2/136}, \href
  {https://ui.adsabs.harvard.edu/abs/2013ApJ...777..136W} {777, 136}

\bibitem[\protect\citeauthoryear{{Wolf} et~al.,}{{Wolf}
  et~al.}{2018}]{skymapper1}
{Wolf} C.,  et~al., 2018, \mn@doi [\pasa] {10.1017/pasa.2018.5}, \href
  {https://ui.adsabs.harvard.edu/abs/2018PASA...35...10W} {35, e010}

\bibitem[\protect\citeauthoryear{{Xu} \& {Li}}{{Xu} \& {Li}}{2010}]{Xu10}
{Xu} X.-J.,  {Li} X.-D.,  2010, \mn@doi [\apj] {10.1088/0004-637X/716/1/114},
  \href {http://adsabs.harvard.edu/abs/2010ApJ...716..114X} {716, 114}

\bibitem[\protect\citeauthoryear{{Yaron}, {Prialnik}, {Shara}  \&
  {Kovetz}}{{Yaron} et~al.}{2005}]{Yaron05}
{Yaron} O.,  {Prialnik} D.,  {Shara} M.~M.,   {Kovetz} A.,  2005, \mn@doi
  [\apj] {10.1086/428435}, \href
  {https://ui.adsabs.harvard.edu/abs/2005ApJ...623..398Y} {623, 398}

\bibitem[\protect\citeauthoryear{{Zahn}}{{Zahn}}{1975}]{Zah75}
{Zahn} J.~P.,  1975, \aap, \href
  {https://ui.adsabs.harvard.edu/abs/1975A&A....41..329Z} {41, 329}

\bibitem[\protect\citeauthoryear{{Zorotovic}, {Schreiber}, {G{\"a}nsicke}  \&
  {Nebot G{\'o}mez-Mor{\'a}n}}{{Zorotovic} et~al.}{2010}]{Zor10}
{Zorotovic} M.,  {Schreiber} M.~R.,  {G{\"a}nsicke} B.~T.,   {Nebot
  G{\'o}mez-Mor{\'a}n} A.,  2010, \mn@doi [\aap] {10.1051/0004-6361/200913658},
  \href {http://adsabs.harvard.edu/abs/2010A%26A...520A..86Z} {520, A86+}

\bibitem[\protect\citeauthoryear{{de Kool}}{{de Kool}}{1990}]{DeK90}
{de Kool} M.,  1990, \mn@doi [\apj] {10.1086/168974}, \href
  {http://adsabs.harvard.edu/abs/1990ApJ...358..189D} {358, 189}

\bibitem[\protect\citeauthoryear{{de Kool}, {van den Heuvel}  \& {Pylyser}}{{de
  Kool} et~al.}{1987}]{DeK87}
{de Kool} M.,  {van den Heuvel} E.~P.~J.,   {Pylyser} E.,  1987, \aap, \href
  {http://adsabs.harvard.edu/abs/1987A%26A...183...47D} {183, 47}

\bibitem[\protect\citeauthoryear{{van Teeseling} \& {King}}{{van Teeseling} \&
  {King}}{1998}]{vanteeseling98}
{van Teeseling} A.,  {King} A.~R.,  1998, \aap, \href
  {https://ui.adsabs.harvard.edu/abs/1998A&A...338..957V} {338, 957}

\bibitem[\protect\citeauthoryear{{van der Helm}, {Portegies Zwart}  \&
  {Pols}}{{van der Helm} et~al.}{2016}]{Van16}
{van der Helm} E.,  {Portegies Zwart} S.,   {Pols} O.,  2016, \mn@doi [\mnras]
  {10.1093/mnras/stv2318}, \href
  {https://ui.adsabs.harvard.edu/abs/2016MNRAS.455..462V} {455, 462}

\bibitem[\protect\citeauthoryear{{von Zeipel}}{{von Zeipel}}{1910}]{Von1910}
{von Zeipel} H.,  1910, \mn@doi [Astronomische Nachrichten]
  {10.1002/asna.19091832202}, \href
  {https://ui.adsabs.harvard.edu/abs/1910AN....183..345V} {183, 345}

\makeatother
\end{thebibliography}

\appendix

\section{The Chance Coincidence Probability}
\label{stats}

In order to assess if the putative proper motion companion to T~Pyx is a chance coincidence, rather than a physically associated object, we have carried out Monte Carlo simulations. We begin by considering the likelihood that an object matching the astrometric properties of T~Pyx as closely as its putative companion should arise purely by chance. We refer to this as the "single-shot" false-alarm probability, since it does not account for the "look-elsewhere effect", i.e. for all of the other "trials" that were carried out as part of the wider search that lead to the discovery of the candidate companion to T~Pyx. We will consider this effect below. 

We estimate the single-shot false-alarm probability by generating 1000 copies of T~Pyx, each with the same parallax, proper motion vector and the associated uncertainties. The only difference between each copy and the true system is their position (i.e. right ascension and declination). The positions of the mock sources around the true source represent the densest packing of 1003 non-overlapping circular search regions\footnote{http://hydra.nat.uni-magdeburg.de/packing/cci/}. With this choice, the sample of {\em Gaia} catalogue sources within each search radius is as statistically similar as possible to, but still completely independent from, the samples within all other search radii. For T~Pyx, our search radius is $\simeq$34\arcsec\  (i.e. the angular radius corresponding to $100,000$~AU at a distance of $\simeq 3.4$~kpc; see above). All of the search regions are then contained with a radius of $\simeq 19.4'$. Even given the position of T~Pyx close to the Galactic plane ($b \simeq +9.7^\circ$), this is small enough to ensure that the distribution of stars within all regions should be statistically similar.

We then search for viable "proper motions companions" to each of the mock sources in exactly the same way as for T~Pyx itself. Of the 1002 mock versions of T~Pyx, only 14 were flagged as having a single viable proper motion companion to within our parallax and proper motion constraints. The single-shot false-alarm probability is therefore $p \simeq 0.014$, i.e. approximately 1\%.

We finally consider the "look-elsewhere" effect. As noted above, T~Pyx was discovered during a search for proper motion companions among $\simeq 1800$ accreting white dwarfs with well-measured parallaxes in {\em Gaia} EDR3. We thus repeated the Monte Carlo analysis described above for T~Pyx for each of these systems. The expected number of chance coincidences across the entire catalogue can then be estimated by summing all of the single-shot false-alarm probabilities. This yields $N_{chance} \simeq 155$.

We can also estimate how many chance coincidences we should expect among sources with single-shot false alarm probabilities as low as T~Pyx, i.e. $p \leq 0.014$. That number is $N_{chance}(p \leq 0.014) \simeq 2$, whereas our actual sample of objects with viable companions contains 23 sources with $p \leq 0.014$, including T~Pyx. Thus 21/23$\simeq 90\%$ of these candidates are likely to be physically associated companions. Of course, the probability of a real association is somewhat higher than this for those sources in this sub-sample with the lowest $p$, and will therefore be somewhat lower for those sources (like T~Pyx), with higher $p$ among this group. We estimate that, for sources with $p \simeq 0.014$, specifically, the probability of a real association is $\simeq 70\%$, in the absence of any other information.

These estimates already show that a real association between T~Pyx and its putative companion is clearly favoured, even taken the look-elsewhere effect into account. We do, however, have additional information: we can check if the properties of T~Pyx's putative companion are statistically consistent with those of the 14 viable (but chance) matches we found among our 1002 displaced copies of T~Pyx. We find that only 1 of these 14 match the target parallax as well as the actual candidate companion, and none match as well in parallax and proper motion jointly. Moreover, 13 of the 14 false matches lie at distances from their targets that are greater than the $\simeq 12$\arcsec\ offset between T~Pyx and its companion. All of this strongly suggests that T~Pyx and its companion are physically related.

\section{Triple Population Synthesis -- Birthrates}
\label{tps}

In order to estimate the birthrates of CVs through channels 1, 2 \& 3, we rely on a triple population synthesis study published previously by one of us \citep{Too20}. We briefly describe the set-up of the simulations below, but refer the reader to the full study for a detailed description and analysis, including visual representations of the initial conditions (their Figs.\,1-3). 

Three populations of triples on the zero-age main-sequence were generated, and their subsequent evolution was simulated for a Hubble time, or until the onset of Roche-lobe overflow, or until the dynamical destabilisation of the system.  We assume an initial triple fraction of 15\%, binary fraction of 50\% and the remainder to be single stars \citep{Tok14b}. To estimate Galactic properties, we assume a constant star formation history of 3 M$_{\odot}$ per year.

The three models differ with respect to their initial distributions of masses and orbital parameters. Model OBin is based on our understanding of observed populations of primordial binaries, whereas model T14 \citep{Tok14b} and model E09 \citep{Egg09} are based on an observed population of triples. We refer to the stars in the inner binary as the primary and secondary with masses $m_1$ and $m_2$, such that initially $m_1>m_2$, and the tertiary has a mass $m_3$. For all models, primary masses are drawn from a Kroupa initial mass function \citep{Kro93} in the range of 1-7.5M$_{\odot}$. The mass ratios of the inner and outer orbit, $q_{\rm in} = m_2/m_1$ and  $q_{\rm out} = m_3/(m_1+m_2)$ respectively, are drawn from a uniform distribution \citep{Rag10,San12,Duc13,Moe17} between 0 and 1 for model OBin and T14. 
For model E09 the inner mass ratio distribution is roughly flat between $0 <q_{\rm in}\leq 1$ except for an enhancement at nearly equal-mass stars \citep{Duc13,Moe17}. The outer mass ratio distribution is also practically flat between $0 <q_{\rm out}\leq 1$ however, with a tail that extends to  $q_{\rm out}> 1$. 
This means that initially the tertiary star can be more massive than the primary in model E09. 

Orbital separations are drawn from a uniform distribution in the log of the orbital separation between $5R_{\odot}$ and $5\times10^6$R$_{\odot}$ for model OBin, a log-normal distribution \citep{Rag10} of periods with $\mu=5$ and $\sigma=2.3$ of for model T14, and by Eggleton's method \citep{Egg09} for model E09. 
Eccentricities are drawn from a thermal distribution between 0 and 1 \citep{Heg75}, 
arguments of pericentre  from a uniform distribution between $-\pi$ and $\pi$, 
and mutual inclinations between the inner and outer orbit from a circular uniform distribution between $0$ and $\pi$ \citep{Tok17}. Initial triples that are dynamically unstable \citep{Mar99, Aar01} are removed from the population. 

\subsection{The Kozai Channel}

From the triple population synthesis, we find that 0.4-0.5\% of all systems will undergo a phase of mass transfer from a MS to a WD without any prior phases of mass transfer. This is not possible for isolated binary evolution. The quoted fraction corresponds to a Galactic birth rate of $(1.9-2.4)\times 10^{-4}$ systems per year. 

In order for a CV to form, we have a number of additional requirements. Firstly, the mass transfer phase needs to proceed in a stable manner, placing an upper limit on the mass of the donor star compared to that of the WD. If we simply exclude donor stars with masses above 1.5M$_{\odot}$, i.e. the canonical upper mass for magnetic braking, the birth rate decreases slightly to $(1.5-1.8)\times 10^{-4}$ systems per year. Furthermore, if we require the donor star to be on the MS, the birth rate is $(0.4-0.8)\times 10^{-4}$ systems per year.
The latter rate should be seen as a lower limit, as \texttt{TRES} underestimates tidal effects at high eccentricities\footnote{
In \texttt{TRES} tides are simulated according to the weak-friction equilibrium-tide model \citep{Zah75, Hut81} in which a tidal bulge is raised that follows the equipotential of the star and that is lagging behind with respect to the companion star. This orbit averaged approach does not hold near to the parabolic regime with small  periastron distances ($5R_{\odot}$), where 
tidal energy is mainly dissipated during the periastron passage  \citep{Fab75,Pre77,Koc92,Mar95,Moe18}. Non-radial dynamical oscillations, which are more efficient in dissipating tidal energy then what is assumed in the simulations presented here, are outside of the scope of this paper.}, 
and does not take into account magnetic braking. If these effects were to be included, systems which currently have a giant donor star would have started mass transfer earlier in the evolution. As such, we estimate the CV formation rate through the Kozai channel to be of the order of $10^{-4}$ events per year.

\subsection{The Dynamical Instability Channel}
\label{sec:app_ch2}

We now consider triples that become dynamical unstable due to their own internal evolution. The fraction of triples that experiences this in our simulated populations is 2.3-4.2\% \citep{Too20,Too21}. This translates to a Galactic event rate of $(1-2)\times 10^{-3}$ per year. After a system crosses the stability limit, it is no longer valid to simulate the evolution of the system with the secular approach as used in \texttt{TRES}. In Toonen et al. subm. \citep{Too21}, we follow up on these systems by simulating their subsequent evolution with the N-body approach using the fourth-order Hermite integrator, while simultaneously including stellar evolution based on the same stellar evolutionary tracks used in \texttt{TRES}. The most common outcome of the dynamically unstable phase is the ejection of one of the stars. This happens in 56-65\% of all destabilized triples. In 21-30\% of all destabilized triples we are left with a WD-MS binary after the ejection. Typically, these binaries are wide (orbital separations $a\sim 10^{4}-10^6 R_{\odot}$), but the distribution of pericenter distances have a long tail to small values. In the extreme, this leads to a collision of two stars, when their passage of closest approach is less than the combination of their stellar radii. In fact, collisions in destabilized triples are common: 13-24\% of all destabilized triples experience a collision, and collisions between a WD and a MS, specifically, occur in 0.7-1.9\% of these systems. The mass of the MS star is typically (in over 90\% of cases) below 1.5M$_{\odot}$. 

As the N-body simulations do not include tidal interactions, nor other dissipative processes, the formation of compact binaries is limited. If tides were included, some of the collisions or ejections would be avoided in the following way. During the dynamically unstable phase, two of the stars of the triple experience a close passage. Tidal dissipation then prevents the stars from carry on along their original trajectories and instead keeps them in a tight orbit around one another. This prevents subsequent close passages in which the stars could collide or become unbound. Close passages of this kind, in which tides can play a role (distances of $\sim 5R_{\odot}$), can also be seen in the evolution of the example system of the dynamical instability channel in Fig.\,\ref{fig:ex2}. 

Including dissipative processes in destabilised triples is outside of the scope of this paper, but the evolution of such systems can, in principle, be followed with a direct N-body code that combines orbital evolution with stellar spins and gravitational tides in a self-consistent manner \citep{TBTIDY}. Based on the calculations we have carried out so far, we expect that the Galactic birth rate via this channel is on the order of $10^{-5}- 10^{-4}$ systems per year. 

\subsection{The Common Envelope Channel}
\label{sec:app_ch3}

To estimate the birthrate of CVs through the CE channel, we focus on all inner binaries that experience a CE-phase. Following classical stability criteria \citep{Hur00}, this entails 54-75\% of all triples. We now need to estimate how many of these systems can avoid a merger of the inner binary and 'survive' the common-envelope phase. To do this, we approximate the post-CE state of the inner binary using the classical $\alpha$-formalism that is based on energy conservation \citep{Pac76,Web84, Liv88, DeK87, DeK90}. According to this framework, orbital energy is consumed to unbind the stellar envelope as:
\begin{equation}
E_{\rm gr} = \alpha (E_{\rm orb,init}-E_{\rm orb,final}),
\label{eq:alpha-ce}
\end{equation}
where $E_{\rm orb}$ is the orbital energy and $E_{\rm gr}$ is the binding energy of the envelope, which is approximated by
\begin{equation}
E_{\rm gr} = \frac{GM_{\rm d} M_{\rm d,env}}{\lambda R}.
\label{eq:Egr}
\end{equation} 
Here, $M_{\rm d}$ is the donor mass, $M_{\rm d, env}$ is the envelope mass of the donor star, $R$ is the radius of the donor star, and $\lambda$ is a structural parameter describing the donor's envelope  \citep{DeK87, Dew00, Xu10, Lov11}. The inclusion of other energy sources \citep{Sok04, Iva13,Nan16, Gla18, Rei20} or other prescriptions  \citep{Nel00, Sok15,Shi19, Law20}  have been discussed extensively in the literature.

The efficiency of orbital energy consumption, expressed by the parameter $\alpha$, has been constrained by studies of post-common envelope \citep{Zor10, Too13,Cam14}. These have found that the CE phase leads to a strong reduction in the orbital separation, i.e. $\alpha\times \lambda\sim 0.25$. This is further confirmed in the massive star regime \citep{Law20}. Adopting this low CE-efficiency, the majority of CE-events lead to a merger, but 10-12\% of all triples form an inner WD-MS binary through a CE-phase. With a less stringent CE-efficiency of $\alpha\times \lambda\sim 1$, as is adopted in classical binary population synthesis studies, the fraction would be 22-28\%.  

To estimate which WD-MS systems will evolve into CVs, we now make two assumptions:
1) orbits will shrink due to gravitational wave emission and magnetic braking \citep{Rap83} until the stars come in contact; 2)  three-body dynamics can be ignored after the CE-phase, i.e. the inner orbit is sufficiently compact to be kinematically decoupled from the outer orbit. We further require that the donor star is still on the MS at the onset of the mass transfer, and that its mass is below 1.5$M_{\odot}$. With these assumptions, a fraction of 6.7-8.0\% of triples would form a CV through a CE-phase. This gives a Galactic birthrate of $(3.3-4.0)\times10^{-3}$ per year. 

As mentioned in the main text, a fundamental difference between triple evolution and binary evolution, is the degree of circularisation at the onset of mass transfer. Even in the CE channel, we find that about 30\% of systems remain significantly eccentric ($e_{\rm in}>0.1$) when the CE develops. Recent hydrodynamical simulations of eccentric CE-phases in binaries \citep{Gla21} demonstrate that initial eccentricities lead to enhanced mass loss, post-CE eccentricities, and smaller post-CE orbital separations compared to the circular case. The post-CE orbital separation of a binary with an initially eccentric orbit ($a_{\rm init}, e_{\rm init}$) is better approximated with that of an initially circular binary with $a=a_{\rm init}(1-e_{\rm init})$ than $a=a_{\rm init}$ \citep{Gla21}. Thus, to first order, the initial pericenter distance determines the post-CE configuration, rather than the initial semi-major axis of the orbit. If we adjust Eq.\,\ref{eq:alpha-ce} accordingly, the CV birthrates do not change in a significant way. 

Any post-CE eccentricity affects the further evolution of the system in two ways. First, it accelerates the orbital shrinkage. For example, a post-CE eccentricity of 0.2 reduces the gravitational wave inspiral time by $\sim$20\% \citep{Gla21}. This does not significantly affect the rates mentioned above either. Second, if the eccentricity is maintained until the onset of the CV-phase, eccentric mass transfer may lead to an enhanced mass transfer rate and kick-start irradiation-induced wind-driven mass transfer  \citep{Knigge00,Too14b}.

\begin{figure}
\includegraphics[width=8.5cm,trim=0cm 0cm 0cm 0.0cm,clip]{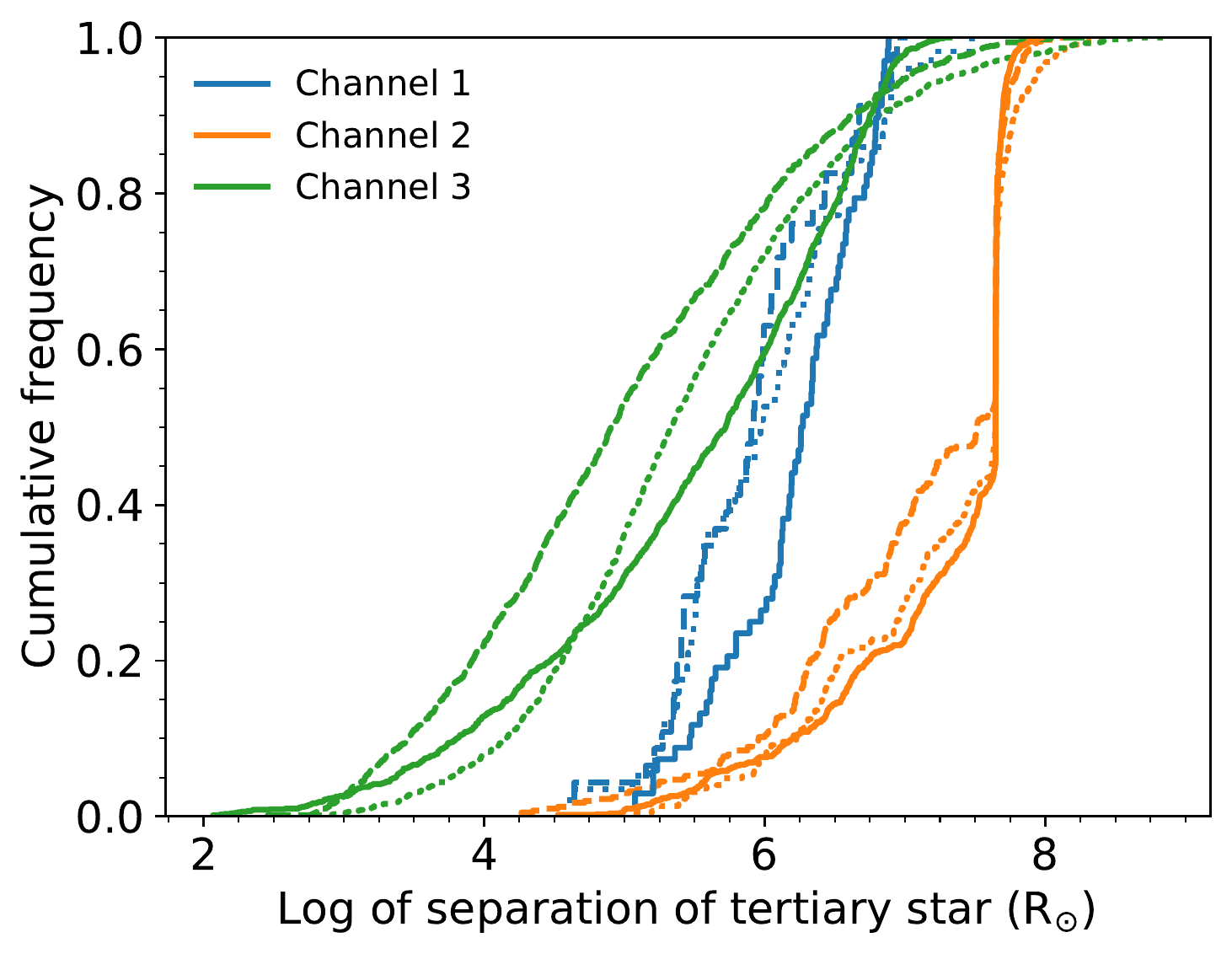} 
\caption{{\bf Distance of the tertiary to the inner binary's centre of mass}  The solid, dashed, and dotted line styles represent the three models for the initial population of triples OBin, T14 and E09, respectively. }
\label{fig:orbit*3}
\end{figure}

\begin{figure*}
    \centering
    \begin{tabular}{cc}
    \includegraphics[width=1\columnwidth]{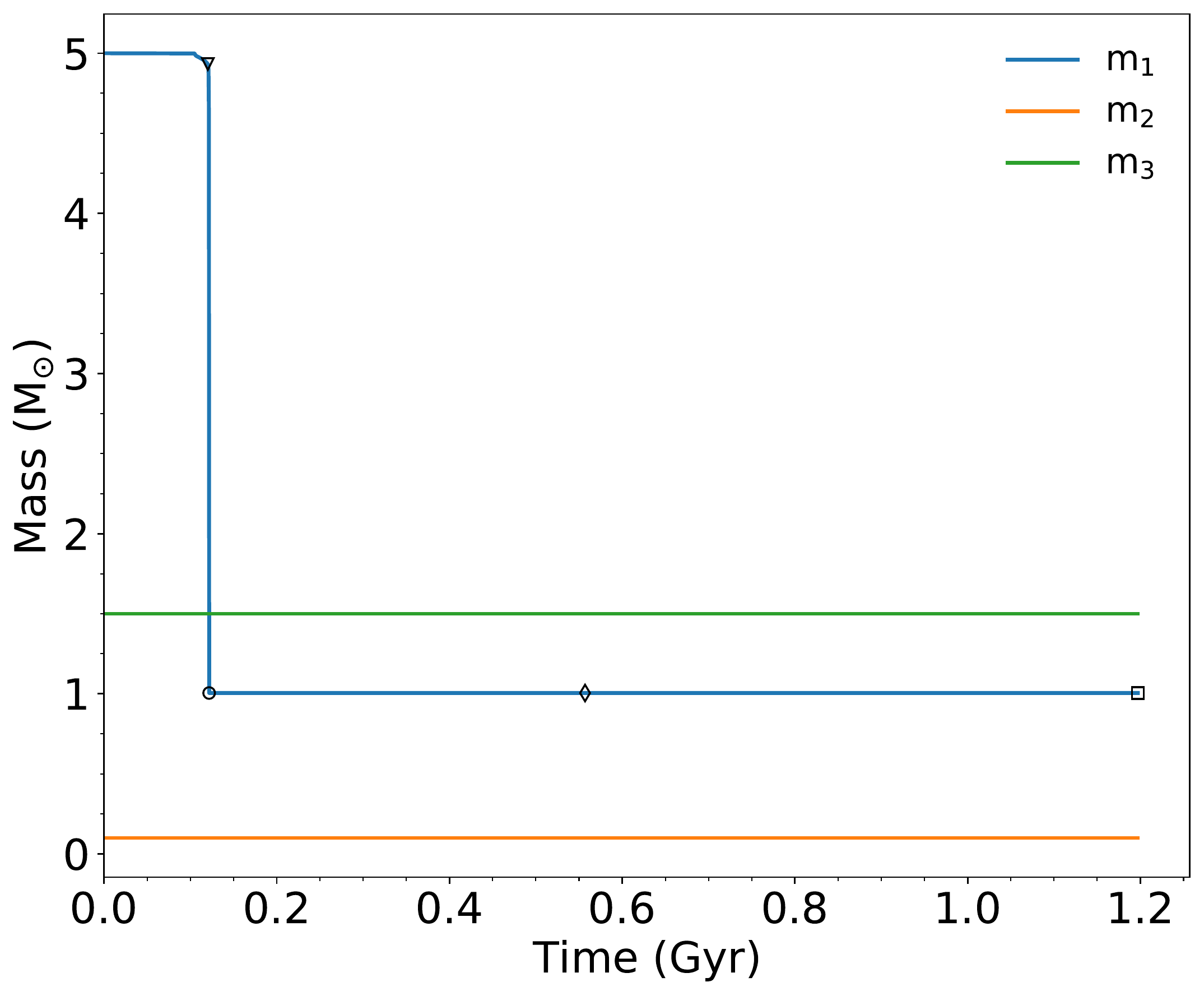}     &  \includegraphics[width=1\columnwidth]{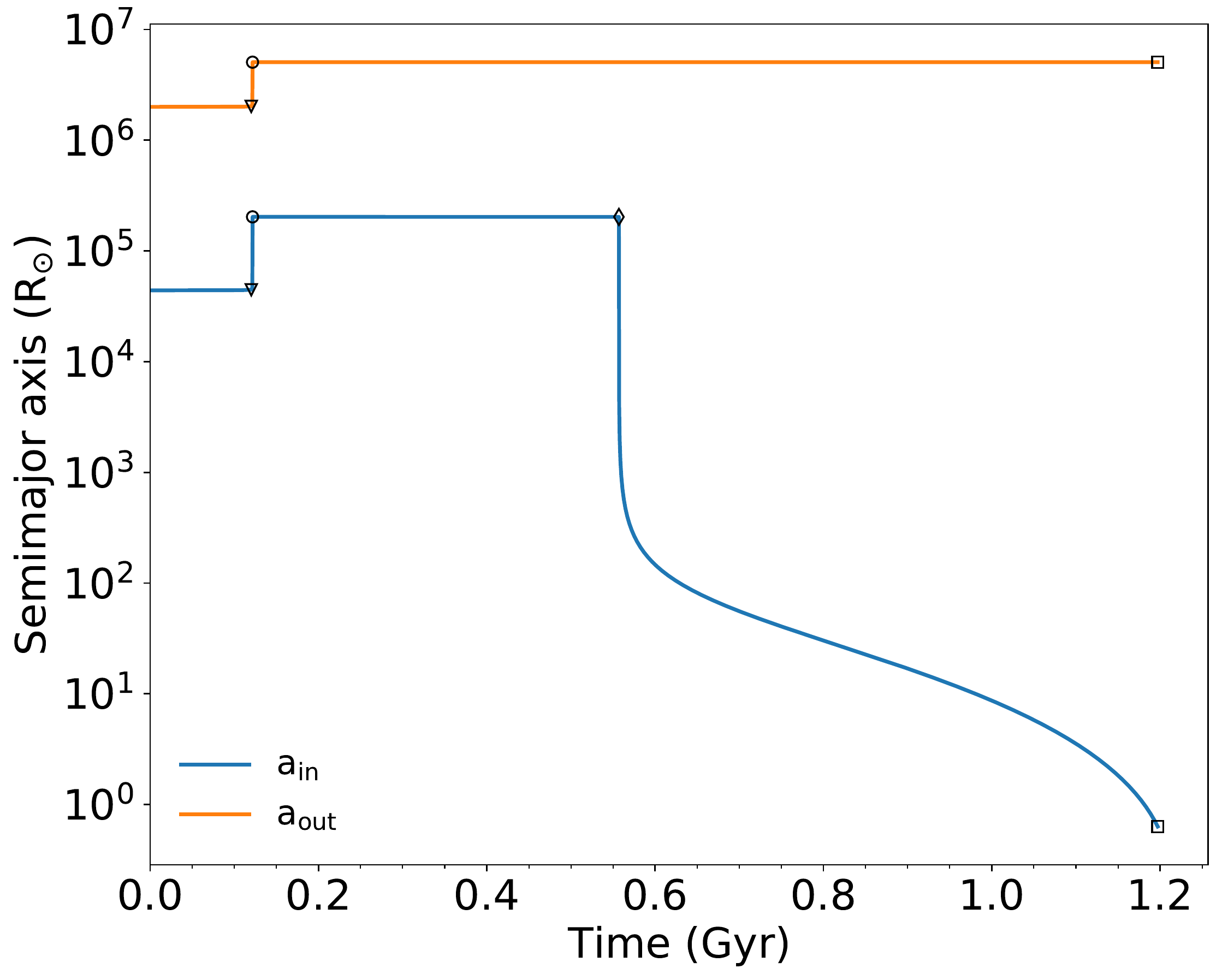} \\
      \includegraphics[width=1\columnwidth]{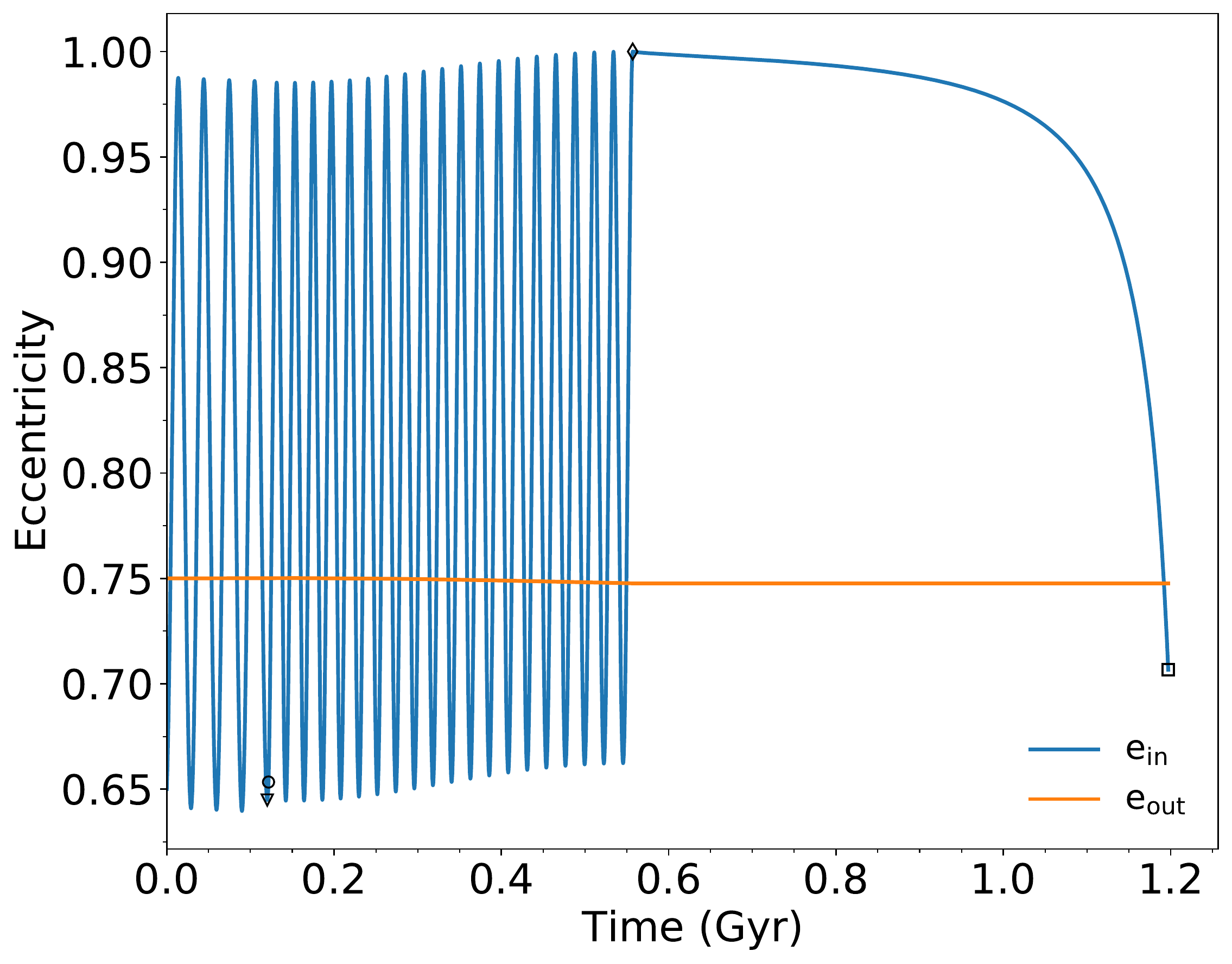}   & 
      \includegraphics[width=1\columnwidth]{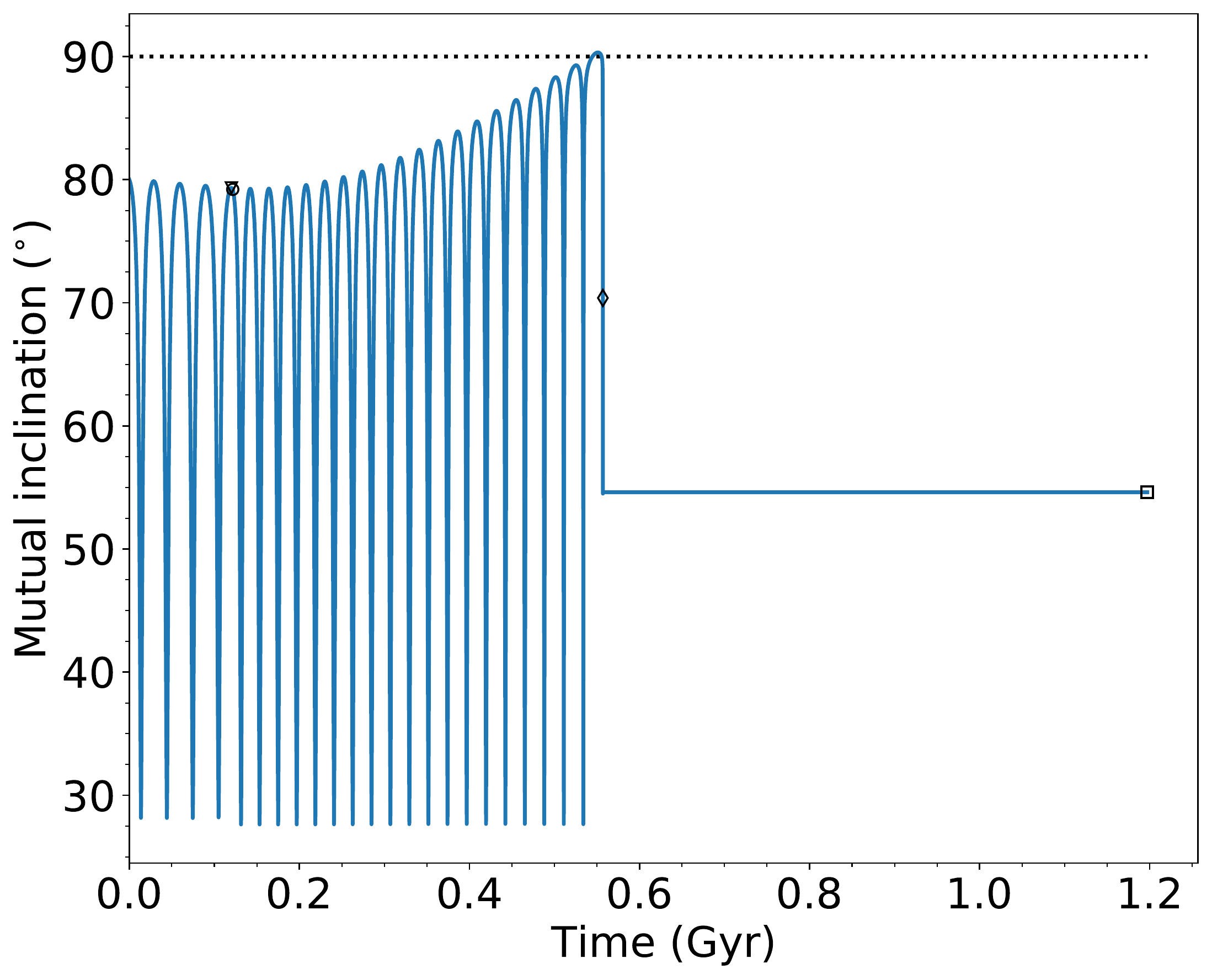} \\
    \end{tabular}
    \caption{\textbf{Example of T-Pyx-like CV formation through the Kozai channel.}
    The primary's arrival on the AGB and WD phases is marked by the triangles and circles, respectively. The onset of the dynamical freeze-out is marked by the diamonds, and the onset of the Roche-lobe overflow (i.e. the birth of the CV) by the squares.   }
    \label{fig:ex1}
\end{figure*}

\begin{figure*}
    \centering
    \begin{tabular}{cc}
      \includegraphics[width=1\columnwidth]{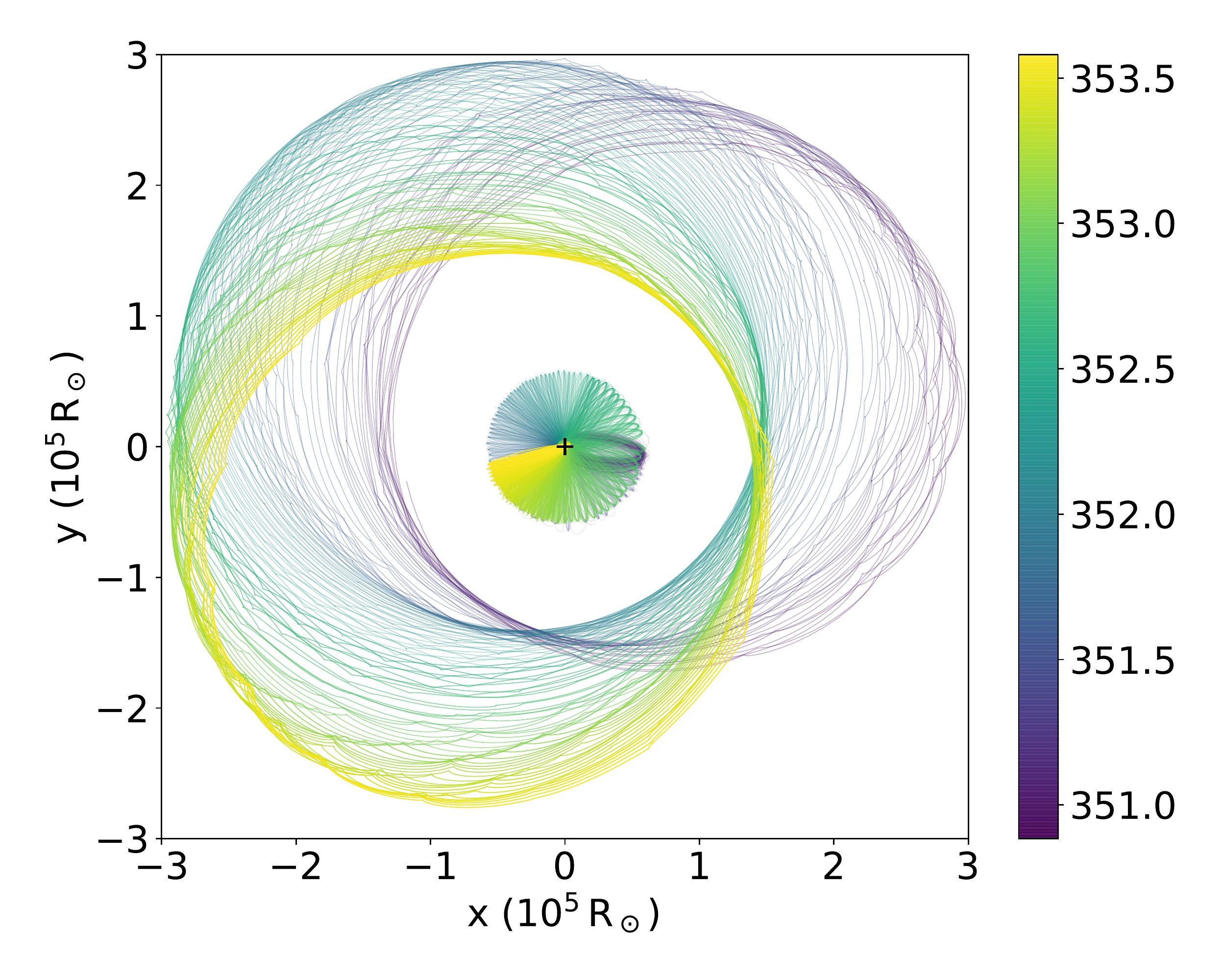}     & 
      
      \includegraphics[width=1\columnwidth]{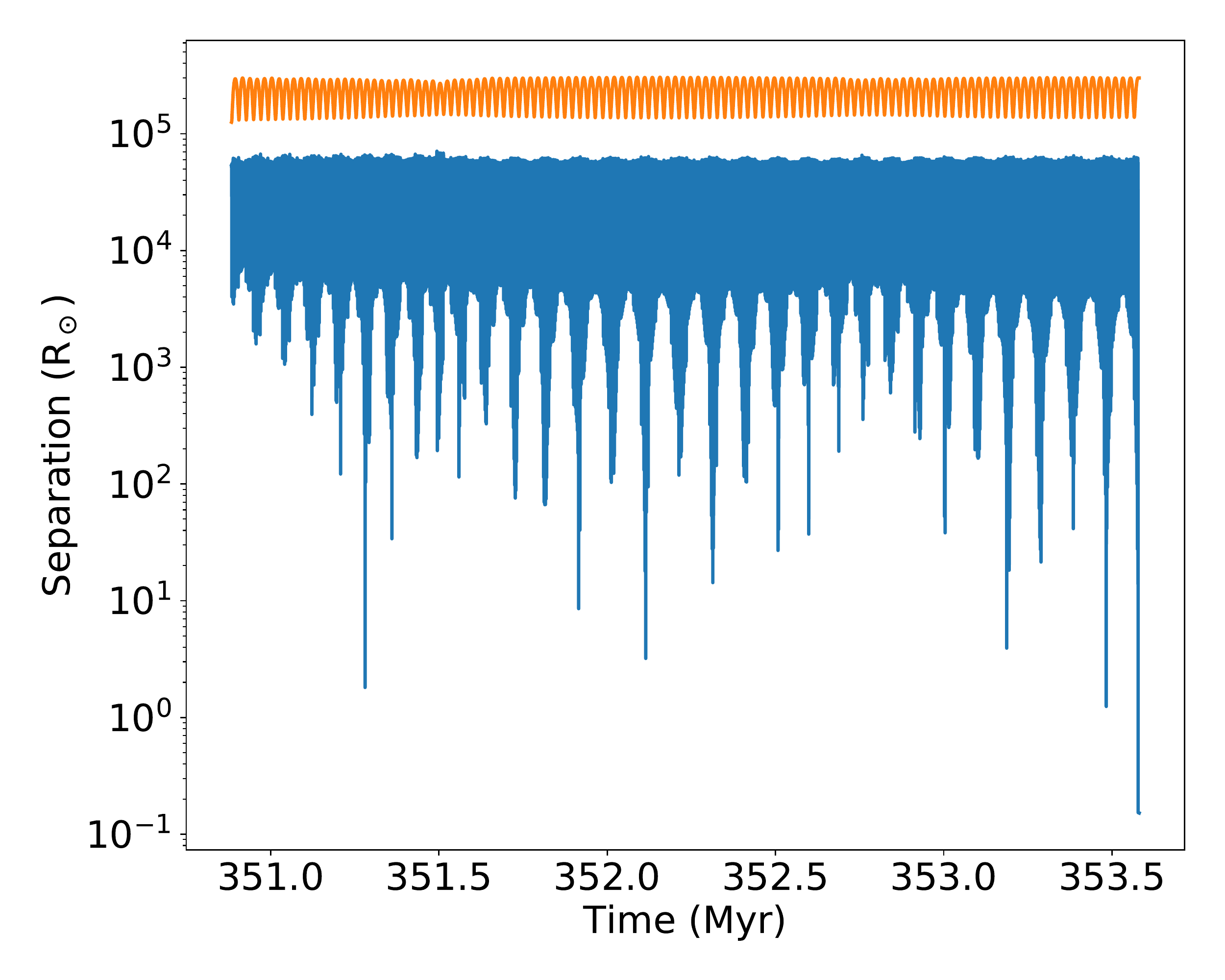}   \\
      \includegraphics[width=1\columnwidth]{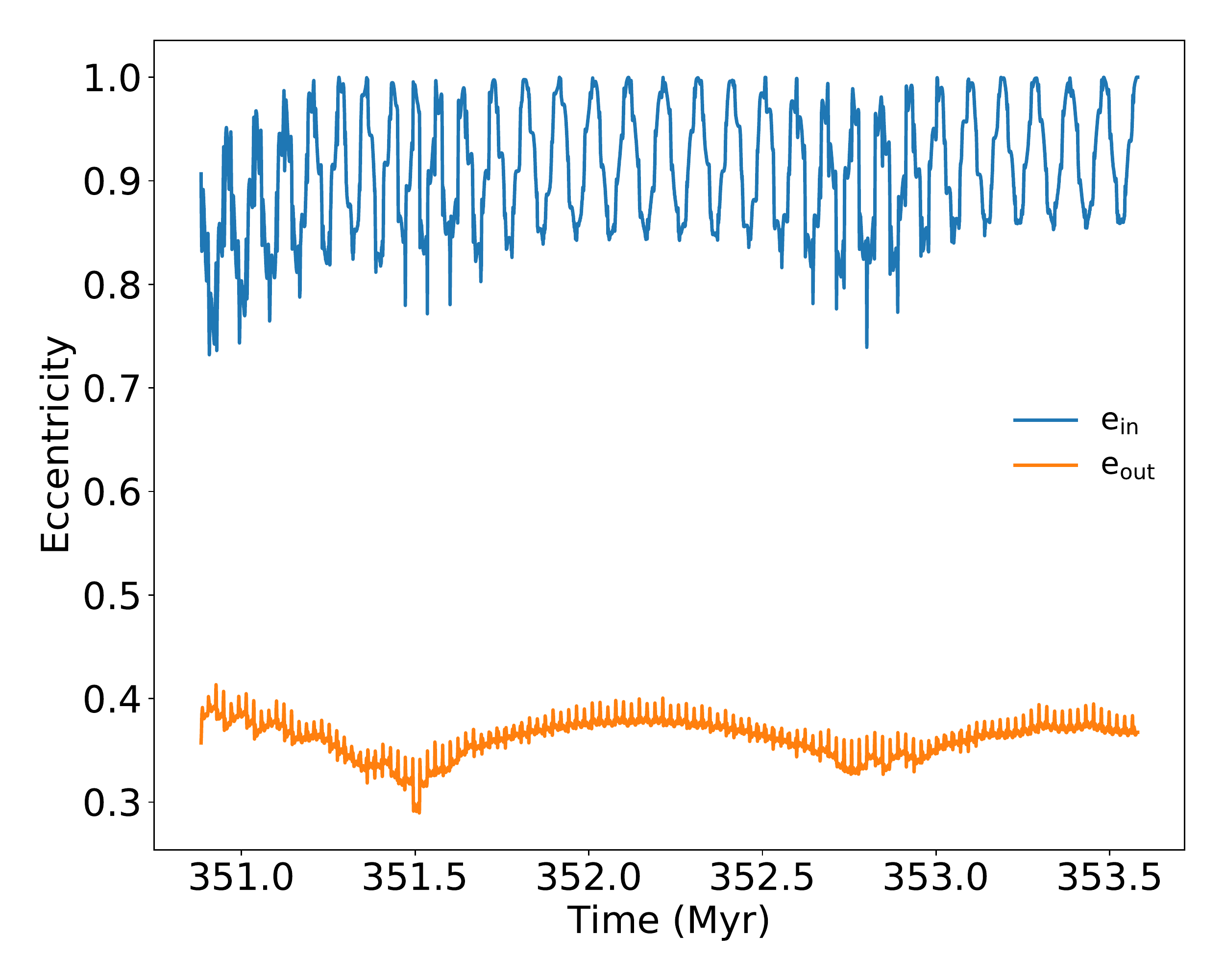} &
       
      \includegraphics[width=1\columnwidth]{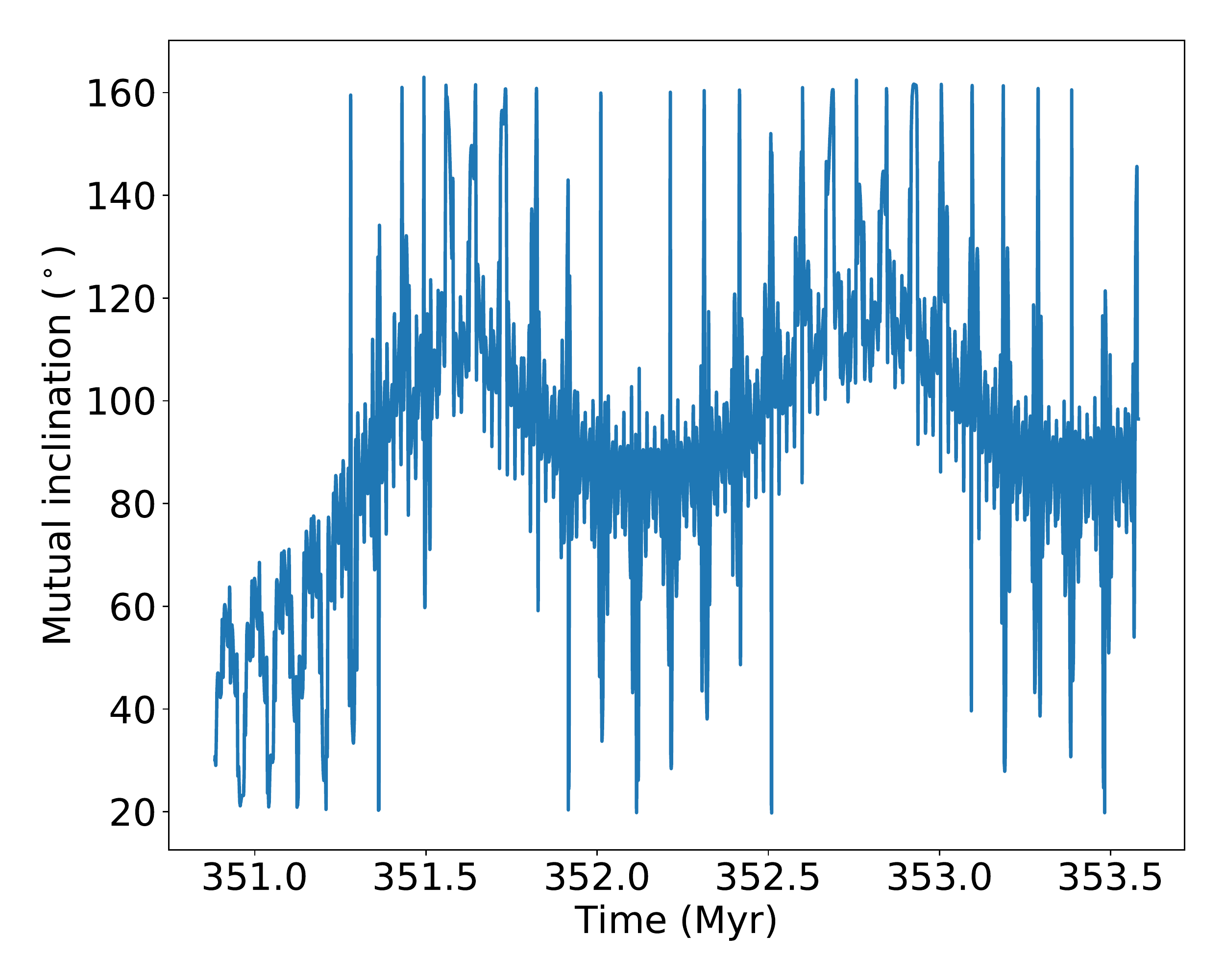} \\
    \end{tabular}
    \caption{
    \textbf{Example of T-Pyx-like CV formation through the dynamical instability channel.}
    At a time of 353.6\,Myr a 
    collision occurs between the WD and the low-mass MS secondary. The top left diagram shows the trajectories of the secondary and tertiary with respect to the WD (indicated by the plus). The color of the tracks presents the time in Myr. 
    The top right shows the distance between the primary and secondary in blue, and
    between the tertiary and center of mass of the inner binary in orange.}
    \label{fig:ex2}
\end{figure*}

\begin{figure*}
    \centering
    \begin{tabular}{cc}
    \includegraphics[width=1\columnwidth]{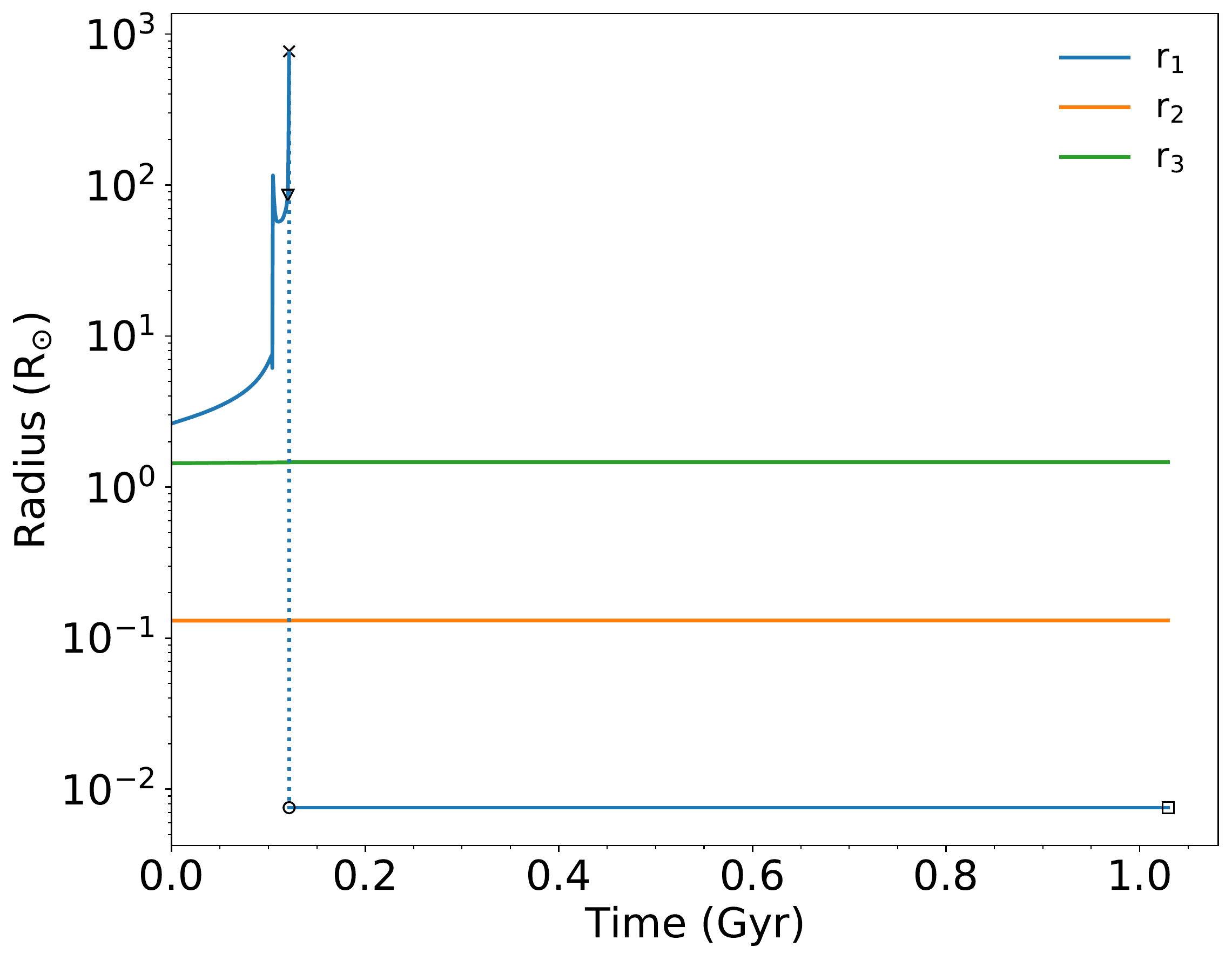}     &  \includegraphics[width=1\columnwidth]{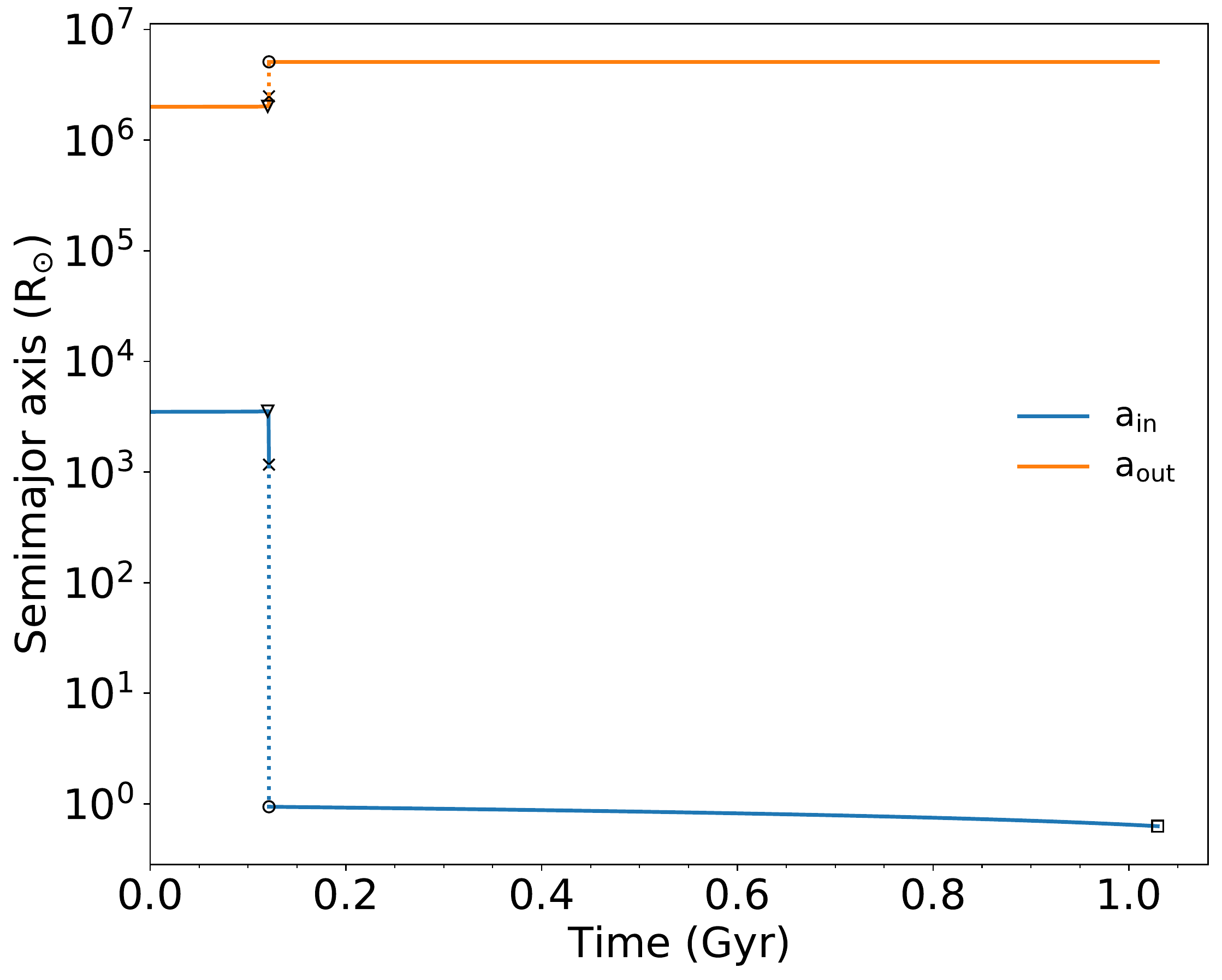} \\
      \includegraphics[width=1\columnwidth]{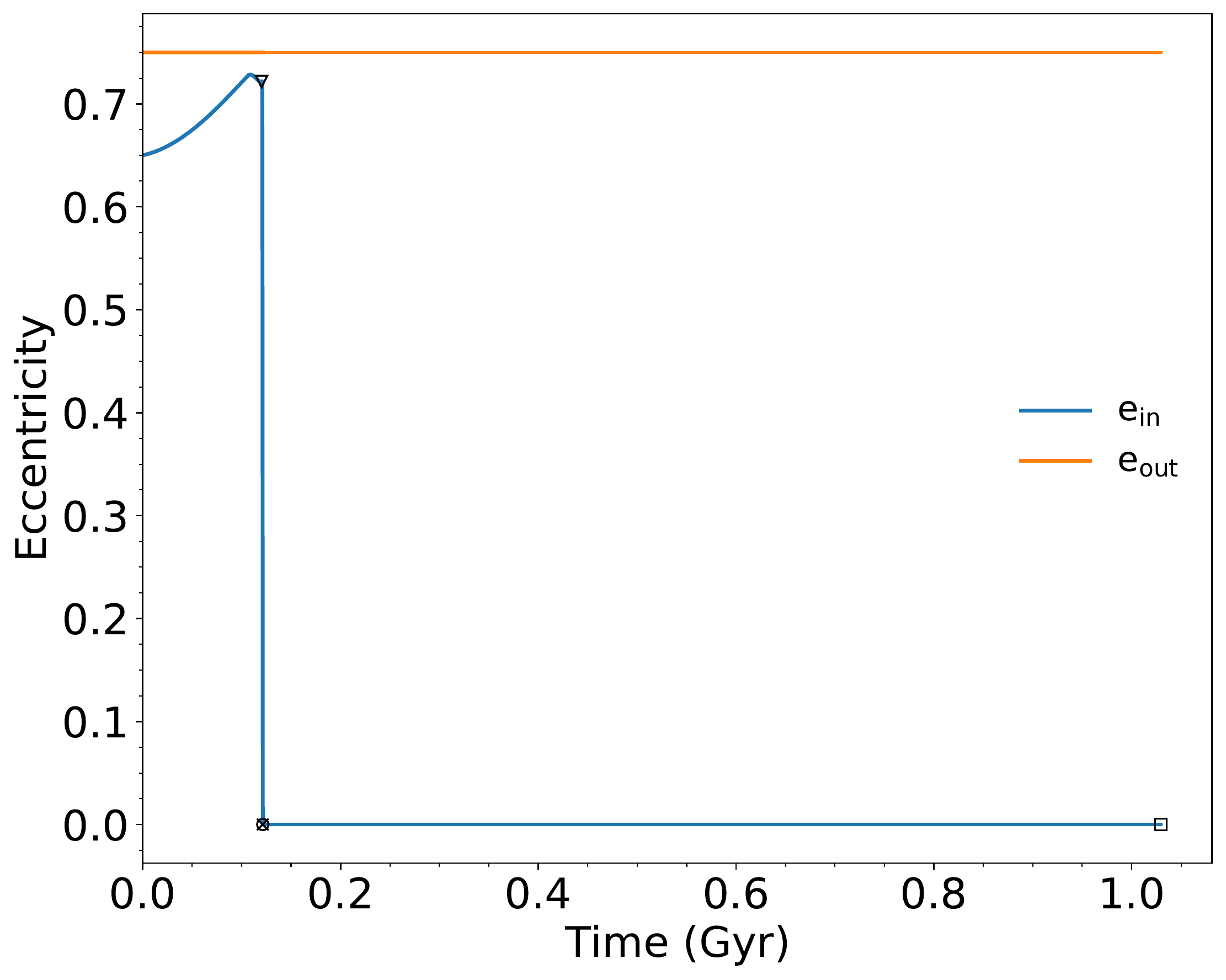}   & 
      \includegraphics[width=1\columnwidth]{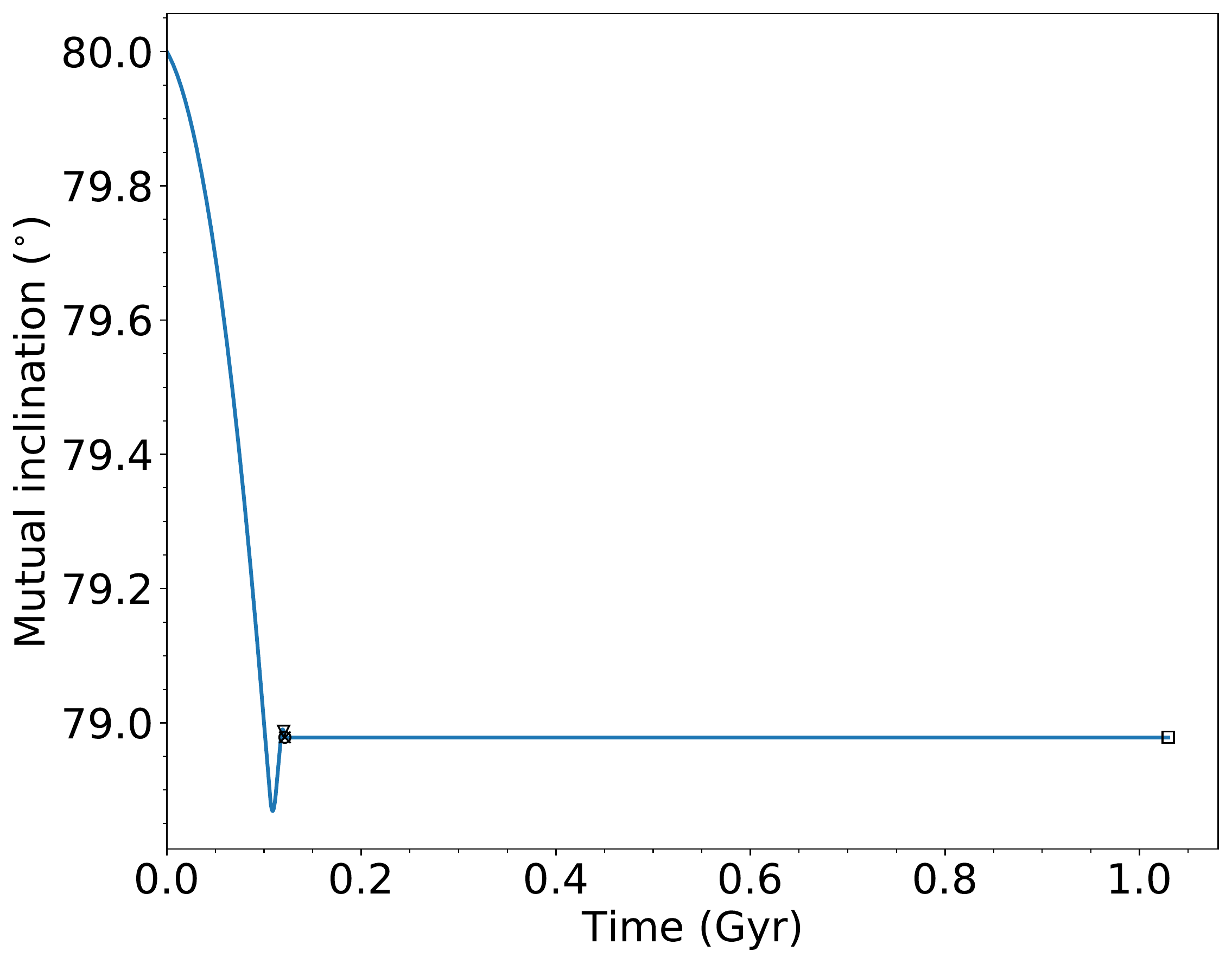} \\
    \end{tabular}
    \caption{\textbf{Example of T-Pyx-like CV formation through channel\,3. }
    The primary's arrival on the AGB and WD phase is marked by the triangles and circles respectively.  The common-envelope phase is marked by the crosses, and the birth of the CV by the squares.   }
    \label{fig:ex3}
\end{figure*}

\section{Triple Population Synthesis -- \\The Orbit of the Tertiary Star}
\label{sec:orbit*3}

For all three channels, the outer star is typically at a large distance from the centre of mass of the inner binary (Fig.\,\ref{fig:orbit*3}). 
Assuming adiabatic winds,  the tertiary remains bound to the inner binary in the Kozai channel. The median orbital separation is $(1-3)\times10^6R_{\odot}$. 
If on the other hand, the mass is lost on a timescale much shorter than the orbital period, the effect is impulsive \citep{Had66,Alc86,Ver11, Too17}. In this case the system experiences a mass-loss kick, similar to that considered in supernovae explosions \citep{Bla61}, and the tertiary star may become unbound from the system.

In the dynamical instability channel, ejection of the tertiary typically occurs at larger orbital distances, $(3-4)\times 10^7R_{\odot}$, with characteristic velocities in the range $0.1-1$km/s. Physical collisions tend to happen in systems with more compact outer orbits, $a_3 \simeq (1.8-3.4)\times 10^5R_{\odot}$. 

The outer orbits in the CE channel are more compact compared to those in the Kozai and dynamical instability channels. At the onset of the CE-phase, they range from about $10^4R_{\odot}$ to about $10^7R_{\odot}$, corresponding to periods of several hundred years to several Myrs. However, the outer orbit will be modified due to the mass loss from the CE-phase in the inner orbit. If the mass is lost in the form of an isotropic wind, and the effect is adiabatic (as usually assumed in binary evolution modelling \citep{Too14}), the outer orbit widens by a factor $m_{\rm tot, pre-CE}/m_{\rm tot, post-CE}\approx 1-2$, where $m_{\rm tot}$ is the total mass of the triple before/after the CE-phase. However, if the mass loss is impulsive, the tertiary star may become unbound from the system \citep{Mic19,Igo20}, as in the case for stellar winds discussed above. 
The timescale of the CE mass ejection, as well as the exact mechanism behind it, is still uncertain \citep{Iva13}.
The dynamical plunge-in phase of the CE, the stage best studied so far, is expected to last several hundreds of years, which is consistent with the observationally derived timescale for CE-evolution in SdB binaries \citep{Igo20}. On the other hand, WD-MS binaries suggest longer CE timescales ($10^3-10^5$yrs) \citep{Mic19, Igo20}, indicating the possible existence of a self-regulating phase after the initial plunge-in \citep{Iva13}. Comparing these timescales to the typical outer orbits of triples undergoing CE-evolution in their inner binaries, we find that some -- or even most -- tertiaries may become unbound due to the CE mass loss. 

\section{Illustrative \texttt{TRES} simulations}

\subsection{A T-Pyx-like CV formed via the Kozai-channel}
The simulation begins with a triple system in which all stars are on the zero-age main-sequence. The inner binary consists of a 5$M_{\odot}$ primary star and a 0.1$M_{\odot}$ secondary, both of which are orbited by a 1.5$M_{\odot}$ tertiary. The orbital separation of the inner binary is $a_{\rm in} =4.4\times10^4R_{\odot}$, while that of the outer orbit is $a_{\rm out} =2.0\times10^6R_{\odot}$. We start the simulation with an inner and outer eccentricity of $e_{\rm in} =0.65$ and $e_{\rm out} =0.75$, which are typical values for wide orbits \citep{Heg75,Duc13,Moe17}. The inner and outer arguments of pericenter are 0.1 and 0.5 radians. The mutual inclination between the orbits is 80$^{\circ}$. 

With these parameters, the system exhibits beautiful Lidov-Kozai cycles (bottom row of Fig.\,\ref{fig:ex1}). Initially, the timescale of these cycles is $\simeq$30 Myr, and the inner eccentricity and inclination vary between 0.64-0.99 and 28-80$^{\circ}$, respectively. After about four cycles, the primary reaches the AGB phase (indicated by the triangles in Fig.\,\ref{fig:ex1}).
The primary loses about 4$M_{\odot}$ of mass in the stellar winds and forms a massive 1$M_{\odot}$ WD (indicated by the circles in Fig.\,\ref{fig:ex1}). The right upper panel of the figure shows that both orbits widen, but the ratio $a_{\rm out}/a_{\rm in}$ decreases - thus the hierarchy of the system decreases. The effect on the triple is immediate. The Lidov-Kozai timescale shortens to $\simeq$22~Myr, but, more importantly, the maximum eccentricity and inclination increase. Eventually, the inclination actually exceeds 90$^{\circ}$, indicating an orbital flip: the system has switched from a prograde orbit to a retrograde one. At this point, the maximum eccentricity is extremely high, but the orbit is wide enough to avoid collisions between the primary and secondary. Instead, strong tidal affects reduce the eccentricity and orbital separation, quenching the Lidov-Kozai cycles (indicated by the diamonds in Fig.\,\ref{fig:ex1}). The mutual inclination eventually freez
es out at $\simeq$55$^{\circ}$. After 1.2 Gyr, when the inner orbital separation has shrunk to 0.6$R_{\odot}$, the 0.1$M_{\odot}$ secondary fills its Roche lobe and starts transferring mass to the WD. The inner eccentricity is still about 0.71 at this point, so mass will be transferred in strong periodic bursts.

\subsection{A T-Pyx-like CV formed via the dynamical instability channel}
If the system described above had formed with a slightly larger inner orbital separation, it would easily have become dynamically unstable during its evolution. For example, with $a_{\rm in}\approx (5-10)\times 10^4 R_{\odot}$ initially,  the triple destabilizes during or shortly after the primary's ascent of the AGB, when it has already developed a large core mass (consistent with the high mass of the WD component in T~Pyx).

In Fig.\,\ref{fig:ex2}, we show the evolution of a system that evolves through such an instability and involves a physical stellar collision. We start with a triple on the zero-age MS with $m_1=3.35M_{\odot}$, $m_2=0.11M_{\odot}$ and $m_3=1.19M_{\odot}$. The initial inner orbit is described by a semimajor axis of $a_{\rm in}=1.1\times 10^4 R_{\odot}$, eccentricity $e_{\rm in} = 0.80$ and an argument of pericenter of $0.93\pi$. For the outer orbit, we adopt $a_{\rm out}=9.5\times 10^4 R_{\odot}$, $e_{\rm in} = 0.36$ and an argument of pericenter of $0.22\pi$. The initial mutual inclination is $55^{\circ}$.

As the system evolves, the primary star loses mass, and the hierarchy of the system reduces. Before the primary's ascent up the AGB, little mass is lost ($\sim 0.03M_{\odot}$). However, during the AGB phase, when the primary mass reaches $0.95M_{\odot}$, and the orbital separations have increased to $a_{\rm in}=2.8\times 10^4 R_{\odot}$ and $a_{\rm out}=19.6\times 10^4R_{\odot}$, the system becomes dynamically unstable \citep{Mar99, Aar01} at 350.9~Myr. 

\texttt{TRES} uses a secular approximation to 3-body dynamics that is not valid once a system becomes dynamically unstable. In this simulation, we therefore switch to an N-body approach based on the fourth-order Hermite integrator, while including radius and mass evolution as in \texttt{TRES} \citep{Too21}. 
Note that tides and gravitational wave emission are not taken into account in this simulation (see also Sect.\,\ref{sec:app_ch2}).
Fig.\,\ref{fig:ex2} shows the subsequent evolution of the orbital trajectories, assuming the mean anomaly of the inner and outer orbits are 180 and 0 degrees, respectively. There are strong variations in eccentricity and inclination on different timescales ($\sim$ 1.2\,Myr and 0.1\,Myr). 
The system experiences many orbital flips, but does not become democratic: the tertiary remains at a large distance from the inner binary, and it does not traverse the inner binary. 
For other combinations of mean anomalies, democratic encounters do occur, even frequently, and lead to the ejection of the very low-mass companion.
In this particular simulation, the eccentricity fluctuations eventually lead to a physical collision during a near-radial encounter after 353.6~Myrs of unstable evolution. 

\subsection{A T-Pyx-like CV formed via the common envelope channel}
Here, we take the same system as in the example for the Kozai channel, but with a more compact inner orbit, $a_{\rm in}=3.5\times10^3R_{\odot}$. During the first 120\,Myr, the system shows evidence of Lidov-Kozai cycling. However, as a result of the increased hierarchy \citep{Kin99,Ant15}, the timescale of the Lidov-Kozai  cycles is now much longer, by about a factor 50. Thus only a portion of a cycle is visible in the bottom panels of Fig.\,\ref{fig:ex3}. 

As the primary's radius expands during its ascent of the AGB, it fills its Roche lobe and initiates a common-envelope phase (indicated by the crosses in Fig.\,\ref{fig:ex3}). 
While the example system for the Kozai channel {\em expands} due to the wind mass loss on the AGB, in Fig.\,\ref{fig:ex3} we see a {\em shrinking} inner orbit. This is due to tides \citep{Maz79, Kis98, Fab07, Liu15, Bat18} and happens despite the stellar wind mass loss. The inner orbit then circularizes in the run-up towards the CE-phase. 

At the onset of the CE-phase, the primary has a mass of 3.67$M_{\odot}$. It then sheds the remainder of its envelope to form a WD of mass 0.99$M_{\odot}$ (indicated by the circles in Fig.\,\ref{fig:ex3}). During the CE-phase, the inner orbit shrinks by three orders of magnitude from $a_{\rm in}\approx 1166R_{\odot}$ to $a_{\rm in}\approx 0.94R_{\odot}$. As the Lidov-Kozai timescale is now of the order of $10^{16}$~yr \citep{Kin99,Ant15}, we can safely ignore three-body dynamical effects after the CE-phase. Over the next 900~Myr, the inner orbit shrinks further due to gravitational wave emission, until the secondary fills its Roche lobe and transfers mass to the WD.

\end{document}